\documentclass[conference]{IEEEtran}
\IEEEoverridecommandlockouts

\usepackage[utf8]{inputenc} 
\usepackage[T1]{fontenc}

\usepackage{booktabs}
\usepackage{listings}
\usepackage[numbers]{natbib}

\newcommand\textcite[1]{\citet{#1}}

\usepackage{adjustbox}
\usepackage{subcaption}
\usepackage{url}

\usepackage{breakurl}
\usepackage[breaklinks]{hyperref}

\usepackage{todonotes}

\usepackage{amsmath,amssymb,amsfonts}
\usepackage{algorithmic}
\usepackage{graphicx}
\usepackage{textcomp}
\usepackage{xcolor}
\usepackage{caption}
\def\BibTeX{{\rm B\kern-.05em{\sc i\kern-.025em b}\kern-.08em
    T\kern-.1667em\lower.7ex\hbox{E}\kern-.125emX}}
\begin{document}

\title{Exploratory Analysis of Covid-19 Tweets using Topic Modeling, UMAP, and DiGraphs}

\author{\IEEEauthorblockN{Catherine Ordun}
\IEEEauthorblockA{Department of Information Systems\\
University of Maryland\\
Baltimore County\\
Booz Allen Hamilton \\
Baltimore, Maryland\\
cordun1@umbc.edu}
\and
\IEEEauthorblockN{Sanjay Purushotham}
\IEEEauthorblockA{Department of Information Systems\\
University of Maryland\\
Baltimore County\\
Baltimore, Maryland\\
psanjay@umbc.edu}
\and
\IEEEauthorblockN{Edward Raff}
\IEEEauthorblockA{Department of Computer Science\\
University of Maryland \\
Baltimore County\\
Booz Allen Hamilton \\
Baltimore, Maryland\\
eraff1@umbc.edu}
}

\maketitle

\begin{abstract}
This paper illustrates five different techniques to assess the distinctiveness of topics, key terms and features, speed of information dissemination, and network behaviors for Covid19 tweets. First, we use pattern matching and second, topic modeling through Latent Dirichlet Allocation (LDA) to generate twenty different topics that discuss case spread, healthcare workers, and personal protective equipment (PPE). One topic specific to U.S. cases would start to uptick immediately after live White House Coronavirus Task Force briefings, implying that many Twitter users are paying attention to government announcements. We contribute machine learning methods not previously reported in the Covid19 Twitter literature. This includes our third method, Uniform Manifold Approximation and Projection (UMAP), that identifies unique clustering-behavior of distinct topics to improve our understanding of important themes in the corpus and help assess the quality of generated topics. Fourth, we calculated retweeting times to understand how fast information about Covid19 propagates on Twitter. Our analysis indicates that the median retweeting time of Covid19 for a sample corpus in March 2020 was 2.87 hours, approximately 50 minutes faster than repostings from Chinese social media about H7N9 in March 2013.  Lastly, we sought to understand retweet cascades, by visualizing the connections of users over time from fast to slow retweeting.  As the time to retweet increases, the density of connections also increase where in our sample, we found distinct users dominating the attention of Covid19 retweeters.  One of the simplest highlights of this analysis is that early-stage descriptive methods like regular expressions can successfully identify high-level themes which were consistently verified as important through every subsequent analysis. 

\end{abstract}


\section{Introduction}
Monitoring public conversations on Twitter about healthcare and policy issues, provides one barometer of American and global sentiment about Covid19. This is particularly valuable as the situation with Covid19 changes every day and is unpredictable during these unprecedented times.  Twitter has been used as an early warning notifier, emergency communication channel, public perception monitor, and proxy public health surveillance data source in a variety of disaster and disease outbreaks from hurricanes\cite{wang2017crisis}, terrorist bombings \cite{buntain2016evaluating}, tsunamis \cite{chatfield2012twitter}, earthquakes \cite{earle2012twitter}, seasonal influenza \cite{nagar2014case}, Swine flu \cite{szomszor2010swineflu}, and Ebola \cite{odlum2015can}. In this paper, we conduct an exploratory analysis of topics and network dynamics of Covid19 tweets.

Since January 2020, there have been a growing number of papers that analyze Twitter activity during the Covid19 pandemic in the United States.  We provide a sample of papers published since January 1, 2020 in \autoref{papers}. Chen, et al. analyzed the frequency of 22 different keywords such as “Coronavirus”, “Corona”, “CDC”, “Wuhan”, “Sinophobia”, and “Covid-19” analyzed across 50 million tweets from January 22, 2020 to March 16, 2020\cite{chen2020covid}. Thelwall also published an analysis of topics for English-language tweets from March 10-29, 2020.\cite{thelwall2020retweeting}. \textcite{singh2020first} analyzed distribution of languages and propogation of myths, Sharma et al. \cite{sharma2020coronavirus} implemented sentiment modeling to understand perception of public policy, and Cinelli et al.\cite{cinelli2020covid} compared Twitter against other social media platforms to model information spread.

\begin{table*}[]
\centering
\caption{Papers published on Covid19 Twitter Analysis since January 2020}
\label{papers}
\resizebox{\textwidth}{!}{%
\begin{tabular}{@{}lllllllllll@{}}
\toprule
Author & Number Tweets & Time Period & Keywords & Feature Analysis & Geospatial & Topic Modeling & Sentiment & Transmission & Network Models & UMAP \\ \midrule
Jahanbin \cite{jahanbin2020using}, et al. & 364,080 & Dec. 31 2019 - Feb. 6 2020 &  &  & x &  &  &  &  &  \\
Banda, et al.\cite{banda2020large} & 30,990,645 & Jan. 1 - Apr 4, 2020 & x &  &  &  &  &  &  &  \\
Medford, et al. \cite{medford2020infodemic} & 126,049 & Jan. 14 - Jan. 28, 2020 & x & x &  & x & x &  &  &  \\
Singh, et al.\cite{singh2020first} & 2,792,513 & Jan. 16, 2020 - Mar. 15, 2020 & x & x & x & x &  &  &  &  \\
Lopez, et al. \cite{lopez2020understanding} & 6,468,526 & Jan. 22 - Mar. 13, 2020 & x & x & x &  &  &  &  &  \\
Cinelli, et al. \cite{cinelli2020covid} & 1,187,482 & Jan. 27 - Feb. 14, 2020 &  & x &  & x &  & x &  &  \\
Kouzy, et al. \cite{kouzy2020coronavirus}& 673 & Feb 27, 2020 & x & x &  &  &  &  &  &  \\
Alshaabi, et al. \cite{alshaabi2020world}& Unknown & Mar. 1 - Mar 21, 2020 & x & x &  &  &  &  &  &  \\
Sharma, et al. \cite{sharma2020coronavirus}& 30,800,000 & Mar. 1, 2020 - Mar. 30, 2020 & x & x & x & x & x & x & x &  \\
Chen, et al. \cite{chen2020covid}& 8,919,411 & Mar. 5, 2020 -  Mar. 12, 2020 & x &  &  &  &  &  &  &  \\
Schild \cite{schild2020go} & 222,212,841 & Nov. 1, 2019 - Mar. 22, 2020 & x & x &  & x &  &  & x &  \\
Yang, et al.\cite{yang2020prevalence} & Unknown & Mar. 9, 2020 - Mar. 29, 2020 & x &  &  &  &  &  & x &  \\
\textbf{Ours} & \textbf{23,830,322} & \textbf{Mar. 24 - Apr. 9, 2020} & \textbf{x} & \textbf{x} & \textbf{} & \textbf{x} & \textbf{} & \textbf{} & \textbf{x} & \textbf{x} \\
Yasin-Kabir, et al.\cite{yasin2020coronavis} & 100,000,000 & Mar. 5, 2020 - Apr. 24, 2020 & x & x & x &  & x &  &  &  \\ \bottomrule
\end{tabular}%
}
\end{table*}

\begin{table*}[]
\centering
\caption{Average Frequency of Keyword Tweets by Minute}
\label{keyword_freq}
\resizebox{\textwidth}{!}{%
\begin{tabular}{@{}llllllllllllll@{}}
\toprule
\textbf{Corpus} &
  \textbf{bed} &
  \textbf{hospital} &
  \textbf{mask} &
  \textbf{icu} &
  \textbf{help} &
  \textbf{nurse} &
  \textbf{doctors} &
  \textbf{vent} &
  \textbf{test\_pos} &
  \textbf{serious\_cond} &
  \textbf{exposure} &
  \textbf{cough} &
  \textbf{fever} \\ \midrule
3/24/2020 & 3.341 & 30.068 & \textbf{38.295} & 3.159 & 2.591 & 4.886 & 8.455  & 25.977 & 0.636 & 0.023 & 0.250 & 0.409 & 0.023 \\
3/25/2020 & 3.117 & 33.021 & \textbf{38.734} & 2.819 & 3.181 & 3.745 & 8.064  & 24.691 & 1.298 & 0.043 & 0.277 & 0.372 & 0.106 \\
3/28/2020 & 1.819 & 30.648 & \textbf{34.352} & 1.714 & 2.362 & 4.800 & 8.486  & 38.790 & 0.962 & 0.019 & 0.181 & 0.181 & 0.029 \\
3/30/2020 & 2.783 & 40.957 & \textbf{53.796} & 2.311 & 3.287 & 6.996 & 13.009 & 24.887 & 1.111 & 0.025 & 0.215 & 0.296 & 0.043 \\
3/31/2020 & 2.109 & 30.673 & \textbf{72.877} & 1.447 & 3.677 & 5.633 & 10.410 & 17.995 & 1.020 & 0.014 & 0.152 & 0.494 & 0.147 \\
4/2/2020  & 2.065 & 29.410 & \textbf{84.467} & 1.474 & 3.164 & 6.147 & 10.450 & 23.424 & 0.814 & 0.018 & 0.192 & 0.357 & 0.045 \\
4/5/2020  & 2.218 & 31.812 & \textbf{62.786} & 2.493 & 3.039 & 5.798 & 10.735 & 17.909 & 1.026 & 0.014 & 0.175 & 0.309 & 0.052 \\
Mean      & 2.493 & 32.370 & \textbf{55.044} & 2.203 & 3.043 & 5.429 & 9.944  & 24.811 & 0.981 & 0.022 & 0.206 & 0.345 & 0.064 \\ \bottomrule
\end{tabular}%
}
\end{table*}

Our contributions are applying machine learning methods not previously analyzed on Covid19 Twitter data, mainly Uniform Manifold Approximation and Projection (UMAP) to visualize LDA generated topics and directed graph visualizations of Covid19 retweet cascades. Topics generated by LDA can be difficult to interpret and while there exist coherence values \cite{roder2015exploring} that are intended to score the interpretability of topics, they continue to be difficult to interpret and are subjective. As a result, we apply UMAP, a dimensionality reduction algorithm and visualization tool that "clusters" documents by topic. Vectorizing the tweets using term-frequency inverse-document-frequency (TF-IDF) and plotting a UMAP visualization with the assigned topics from LDA allowed us to identify strongly localized and distinct topics. We then visualized  "retweet cascades", which describes how a social media network propagates information \cite{jin2017detection}, through the use of graph models to understand how dense networks become over time and which users dominate the Covid19 conversations.  

In our retweeting time analysis, we found that the median time for Covid19 messages to be retweeted is approximately 50 minutes faster than H7N9 messages during a March 2013 outbreak in China, possibly indicating the global nature, volume, and intensity of the Covid19 pandemic.  Our keyword analysis and topic modeling were also rigorously explored, where we found that specific topics were triggered to uptick by Live White House Briefings, implying that Covid19 Twitter users are highly attuned to government broadcasts. We think this is important because it highlights how other researchers have identified that government agencies play a critical role in sharing information via Twitter to improve situational awareness and disaster response \cite{genes2014analysis}.  Our LDA models confirm that topics detected by Thelwall et al. \cite{thelwall2020retweeting} and Sharma et al. \cite{sharma2020coronavirus}, who analyzed Twitter during a similar period of time, were also identified in our dataset which emphasized healthcare providers, personal protective equipment such as masks and ventilators, and cases of death.  

\subsection{Research Questions} 
This paper studies five research questions: 
\begin{enumerate}
\item What high-level trends can be inferred from Covid19 tweets?
\item Are there any events that lead to spikes in Covid19 Twitter activity?
\item Which topics are distinct from each other? 
\item How does the speed of retweeting in Covid19 compare to other emergencies, and especially similar infectious disease outbreaks?
\item How do Covid19 networks behave as information spreads?
\end{enumerate}

The paper begins with Data Collection, followed by the five stages of our analysis: Keyword Trend Analysis, Topic Modeling, UMAP, Time-to-Retweet Analysis, and Network Analysis. Our methods and results are explained in each section. The paper concludes with limitations of our analysis. The Appendix provides additional graphs as supporting evidence.

\section{Data Collection}
Similar to researchers in \autoref{papers}, we collected Twitter data by leveraging the free Streaming API. From March 24, 2020 to April 9, 2020, we collected 23,830,322 (173 GB) tweets. Note, in this paper, we refer to the Twitter data interchangeably as both "dataset" and "corpora" and refer to the posts as "tweets". Our dataset is a collection of tweets from different time periods shown in \autoref{datasets}. Using the Twitter API through tweepy, a Python Twitter mining and authentication API, we first queried the Twitter track on twelve query terms to capture a healthcare-focused dataset:  'ICU beds', 'ppe', 'masks', 'long hours', 'deaths', 'hospitalized', 'cases', 'ventilators', 'respiratory', 'hospitals', '\#covid', and '\#coronavirus'.  For the keyword analysis, topic modeling, and UMAP tasks, we analyzed non-retweets that brought the corpus down to 5,506,223 tweets. In the Time-to-Retweet and Network Analysis, we included retweets but selected a sample out of the larger 23.8 million corpus of 736,561 tweets. Our preprocessing steps are described in the Data Analysis section that follows.

\section{Keyword Trend Analysis}\label{AA}
Prior to applying keyword analysis, we first had to pre-process the corpus on the “text” field. First, we removed retweets using regular expressions, in order to focus the text on original tweets and authorship, as opposed to retweets that can inflate the number of messages in the corpus. We use no-retweeted corpora for both the keyword trend analysis and the topic modeling and UMAP analyses. Further we formatted datetime to UTC format, removed digits, short words less than 3 characters, extended the NLTK stopwords list to also exclude “coronavirus”, “covid19”, “19”, “covid", removed “https:” hyperlinks, removed “@” signs for usernames, removed non-Latin characters such as Arabic or Chinese characters, and implemented lower-casing, stemming, and tokenization. Finally, using  regular expressions, we extracted tweets that contained the following thirteen single terms: 'bed', 'hospital',  'mask', 'icu', 'help',  'nurse', 'doctors', 'vent', 'test\_pos’,  'serious\_cond', 'exposure', 'cough', and 'fever', in order to gain insights about currently trending public concerns. We present values of the raw counts of the tweets in the Appendix under \autoref{appendix_rawcounts} and the frequencies of tweets per minute here in \autoref{keyword_freq}.

The greatest rate of tweets occurred for the tweets consisting of the term "mask" (mean 55.044) in \autoref{keyword_freq}, followed by "hospital" (mean 32.370)  and  "vent" (mean 24.811). Tweets of less than 1.0 mean tweets per minute, came from groups about testing positive, being in serious condition, exposure, cough, and fever.  This may indicate that people are discussing the issues around Covid19 more frequently than symptoms and health conditions in this dataset. We will later find out that several themes consistent with these keyword findings are mentioned in topic modeling to include personal protective equipment (PPE) like ventilators and masks, and healthcare workers like nurses and doctors. 

\section{Topic Modeling}
LDA are mixture models, meaning that documents can belong to multiple topics and membership is fractional \cite{blei2003latent}. Further, each topic is a mixture of words, where words can be shared among topics. This allows for a "fuzzy" form of unsupervised clustering where a single document can belong to multiple topics, each with an associated probability. LDA is a bag of words model where each vector is a count of terms. LDA requires the number of topics to be specified. Similar to methods described by Syed et al. \cite{syed2017full}, we ran 15 different LDA experiments varying the number of topics from 2 to 30, and selected the model with the highest coherence value score. We selected the LDA model that generated 20 topics, with a medium coherence value score of 0.344. Roder et al. \cite{roder2015exploring} developed the coherence value as a metric that calculates the agreement of a set of pairs and word subsets and their associated word probabilities into a single score. In general, topics are interpreted as being coherent if all or most of terms are related. 

\begin{figure}[t!]
\centering
\includegraphics[width=\columnwidth]{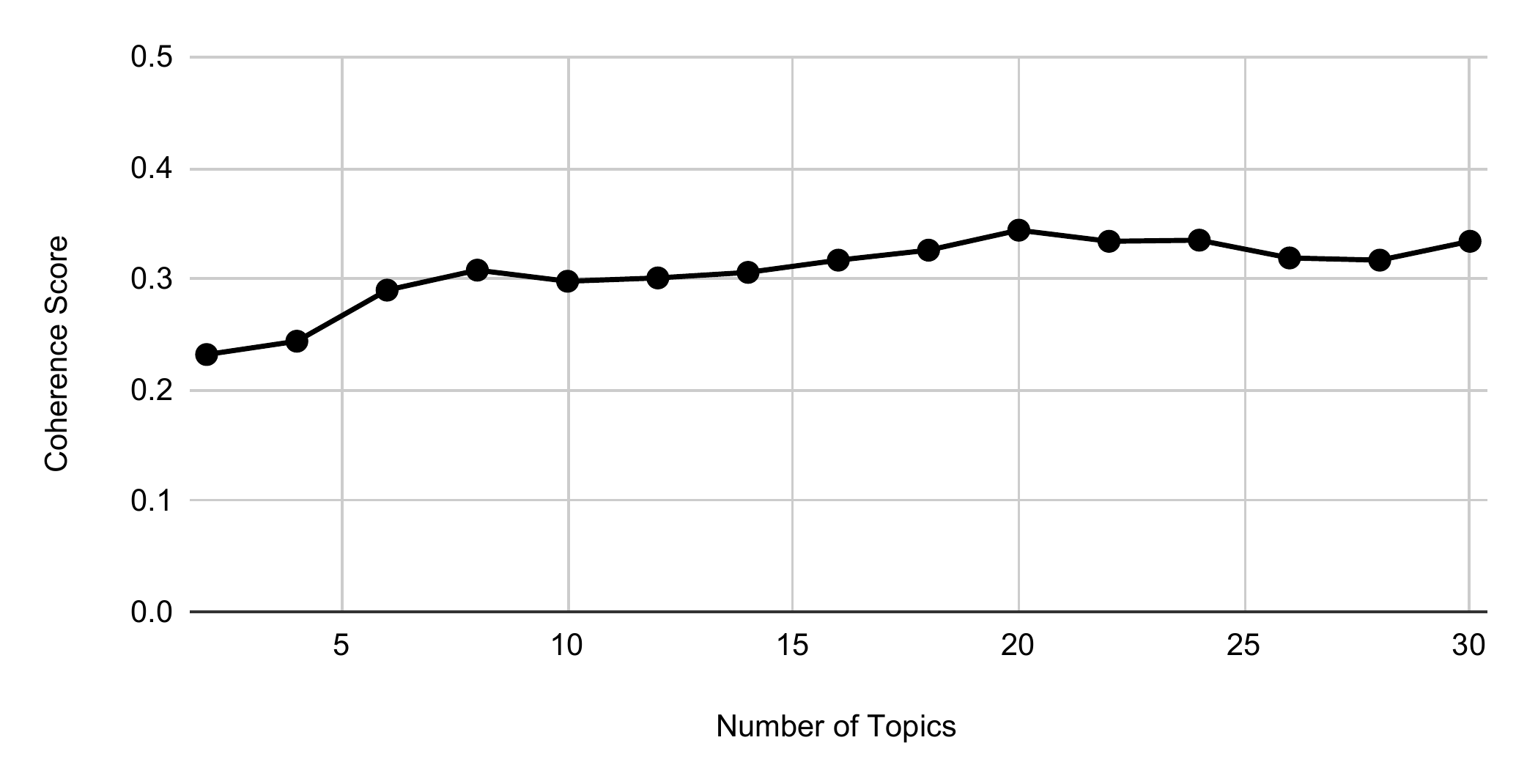}
\caption{Coherence Scores by Number of Topics}
\label{topics_dist}
\end{figure}


\begin{table*}[]
\centering
\caption{20 Topics Generated from LDA Model}
\label{topics}
\resizebox{\textwidth}{!}{%
\begin{tabular}{@{}llll@{}}
\toprule
Topic & C\_V  & Terms                                                                                                                                        & Language   \\ \midrule
1     & 0.922 & de, la, el, en, que, lo, por, del, para, se, es, con, un, al, est, una, su, ms, caso, todo                                                   & Spanish    \\
2     & 0.241 & like, look, work, dont, amp, peopl, time, read, support, respiratori, great, death, us, case, hospit, listen, im, presid, agre, way          & English    \\
3     & 0.222 & hospit, realli, patient, johnson, bori, oh, shit, amp, peopl, make, death, e, blood, like, call, treat, human, trial, guy                    & English    \\
4     & 0.171 & china, thank, lockdown, viru, latest, corona, pandem, covid2019, us, lie, hai, ye, stayhom, trump, daili, way, social, quarantin, help, 5g   & English    \\
5     & 0.363 & case, spread, help, slow, risk, symptom, daili, mask, identifi, sooner, asymptomat, us, test, market, selfreport, de, 2, 9, question, commun & English    \\
6     & 0.413 & day, case, week, news, ago, state, health, two, month, death, last, 15, us, delhi, hospit, one, 2, new, said, lockdown                       & English    \\
7     & 0.287 & test, case, hospit, posit, corona, dr, viru, kit, patient, ppe, doctor, data, govern, work, de, say, vaccin, death, drug, amp                & English    \\
8     & 0.173 & die, world, peopl, case, us, death, der, tell, und, flu, corona, da, im, never, cant, fr, thousand, africa, help, ist                        & English    \\
9     & 0.413 & mask, face, wear, make, one, public, protect, cdc, peopl, dont, n95, recommend, us, viru, love, cloth, new, 0, trump, work                   & English    \\
10    & 0.440 & mask, home, stay, peopl, pleas, ppe, hospit, help, work, wear, amp, like, worker, care, nurs, safe, sure, dont, doctor, hand                 & English    \\
11    & 0.296 & hospit, nurs, le, case, de, ppe, work, new, doctor, go, pay, help, let, one, live, us, local, time, staff, lockdown                          & English    \\
12    & 0.572 & case, death, new, report, total, confirm, day, posit, number, york, us, state, 1, today, 2, 3, updat, test, peopl, rise                      & English    \\
13    & 0.483 & mask, ppe, ventil, hospit, medic, trump, suppli, donat, us, need, worker, state, china, n95, million, use, help, order, equip, amp           & English    \\
14    & 0.713 & de, que, e, em, da, per, el, com, la, para, um, se, os, le, na, un, mai, brasil, dia, del                                                    & Portuguese \\
15    & 0.490 & case, death, number, total, countri, updat, time, india, confirm, recov, china, corona, hour, last, us, news, peopl, new, activ, hospit      & English    \\
16    & 0.582 & di, il, e, la, na, per, che, non, sa, al, si, un, da, del, ng, ang, le, ha, con, het                                                         & Italian    \\
17    & 0.247 & great, god, news, sad, shame, ppe, bless, hydroxychloroquin, hospit, de, death, ventil, stori, die, amp, hear, man, case, hong, holi         & English    \\
18    & 0.329 & trump, peopl, death, american, live, stop, amp, us, let, hospit, time, viru, caus, like, one, dont, true, go, kill, media                    & English    \\
19    & 0.904 & de, le la, en, et, du, pour, un, pa, que, il, ce, au, qui, confin, dan, une, est, cest, sur                                                  & French     \\
20    & 0.293 & hospit, im, peopl, still, govern, dont, thing, amp, death, fuck, one, work, job, state, money, model, us, start, happen, ive                 & English    \\ \bottomrule
\end{tabular}%
}
\end{table*}

\begin{figure}[t!]
\centering
\includegraphics[width=\columnwidth]{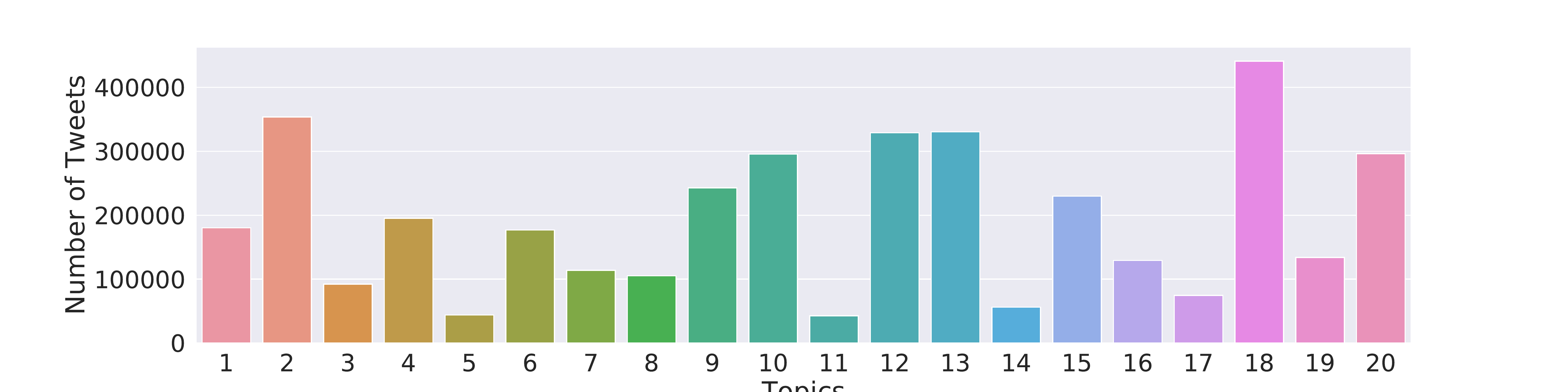}
\caption{Distribution of 20 Topics in the Corpora}
\label{topics_dist}
\end{figure}

\begin{figure*}[t!]
\centering
\includegraphics[width=1.0\textwidth]{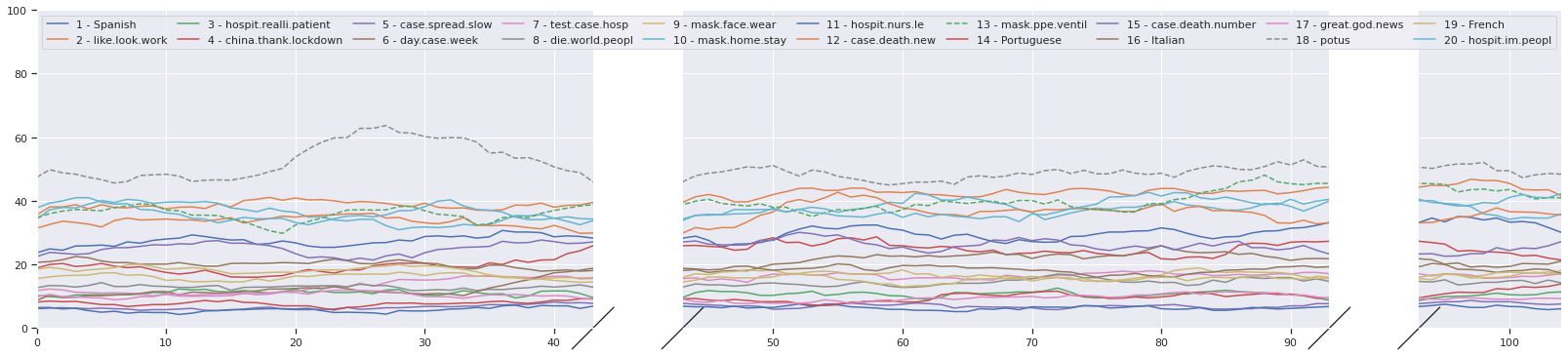}
\caption{Trend of Topics over Time from March 24 to March 28, 2020}
\label{baxes1}
\end{figure*}

\begin{figure*}[t!]
\centering
\includegraphics[width=1.0\textwidth]{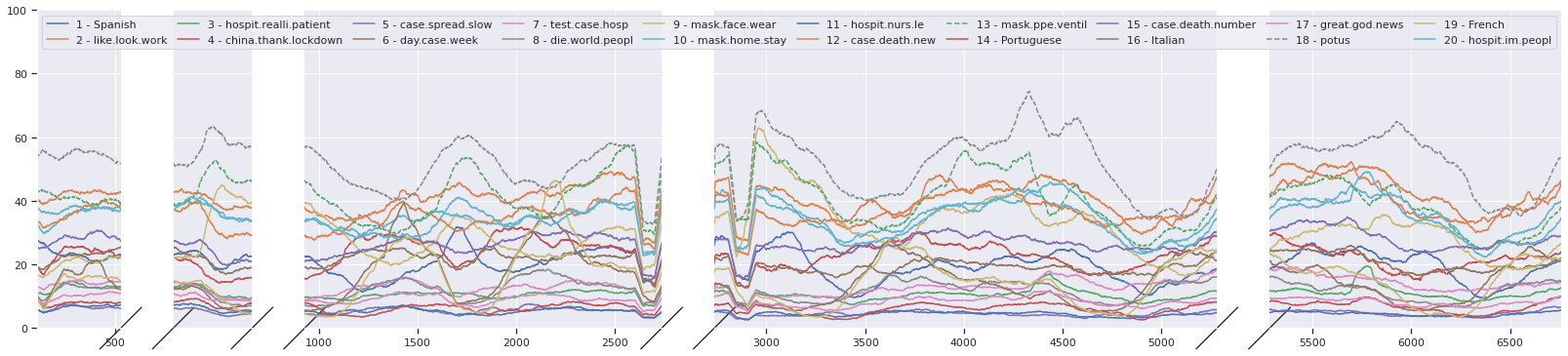}
\caption{Trend of Topics over Time from March 30 to April 8, 2020}
\label{baxes2}
\end{figure*}

\begin{figure*}[ht]
\centering
\includegraphics[width=1.0\textwidth]{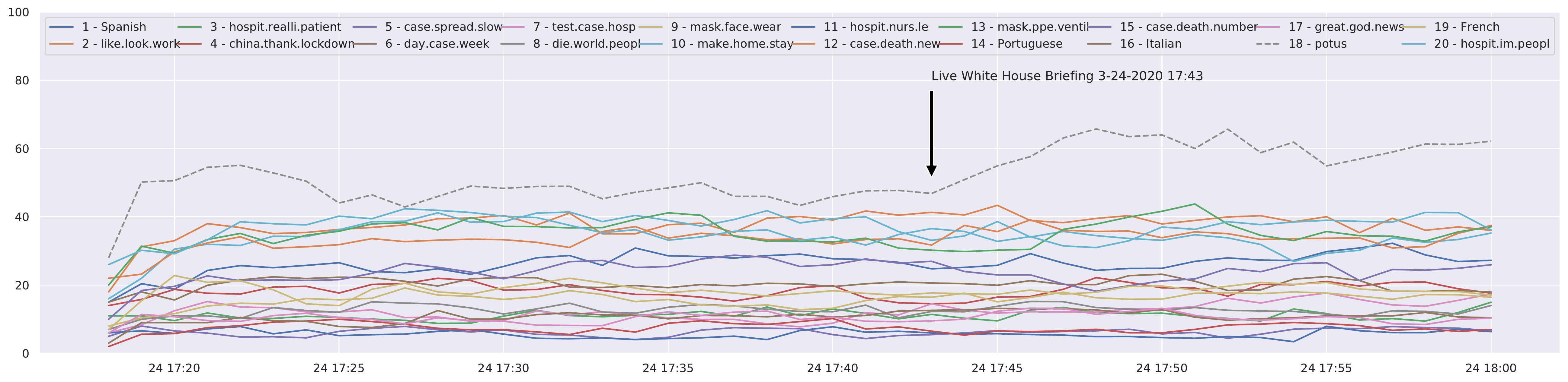}
\caption{March 24 5:17 PM to 6:00 PM EST Topics Time Series}
\label{WHbriefing}
\end{figure*}

\begin{figure*}[ht]
\centering
\includegraphics[width=0.8\textwidth]{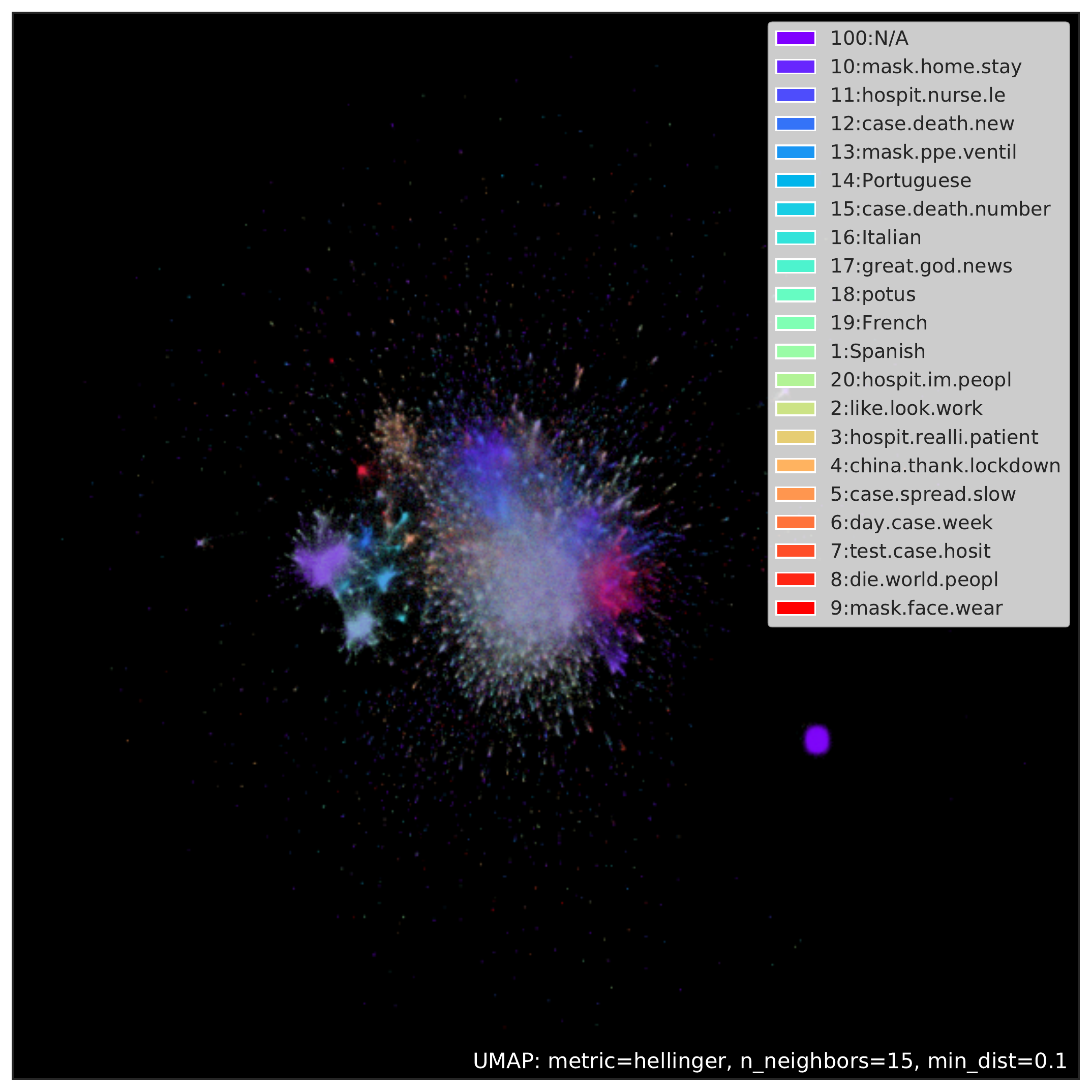}
\caption{Visualization of One Million Tweets with Topic Labels}
\label{umap}
\end{figure*}


Our final model generated 20 topics using the default parameters of the Gensim LDA MultiCore model \footnote{https://radimrehurek.com/gensim/models/ldamulticore.html} with an overall coherence score of 0.428 after modifying the chunksize to 50,000. The topics are provided in \autoref{topics_dist} and include the terms generated and each topic's coherence score measuring interpretability. Similar to the high-level trends inferred from extracting keywords, themes about PPE and healthcare workers dominate the nature of topics. The terms generated also indicate emerging words in public conversation including "hydroxychloroquine" and "asymptomatic". 

Our results also show four topics that are in non-English languages. In our preprocessing, we removed non-Latin characters in order to filter out a high volume of Arabic and Chinese characters. In Twitter there exists a Tweet object metadata field of "lang" for language to filter tweets by a specific language like English ("eng"). However, we decided not to filter against the "lang" element because upon observation, approximately 2.5\% of the dataset consisted of an "undefined" language tag, meaning that no language was indicated. Although it appears to be a small fraction, removing even the "undefined" tweets would have removed several thousand tweets. Some of these tweets that are tagged as "undefined" are in English but contain hashtags, emojis, and Arabic characters. As a result, we did not filter out for English language, leading our topics to be a mix of English, Spanish, Italian, French, and Portuguese. Although this introduced challenges in interpretation, we feel it demonstrates the global nature of worldwide conversations about Covid19 occurring on Twitter. This is consistent with what Singh et al. \textcite{singh2020first} reported as a variety of languages in Covid19 tweets upon analyzing over 2 million tweets. As a result, we labeled the four topics by the language of the terms in the respective topics: "Spanish" (Topic 1), "Portuguese" (Topic 14), "Italian" (Topic 16) and "French" (Topic 19). We used Google Translate to infer the language of the terms.


\begin{figure*}
\begin{subfigure}{\textwidth}
\centering
  \includegraphics[width=0.75\linewidth]{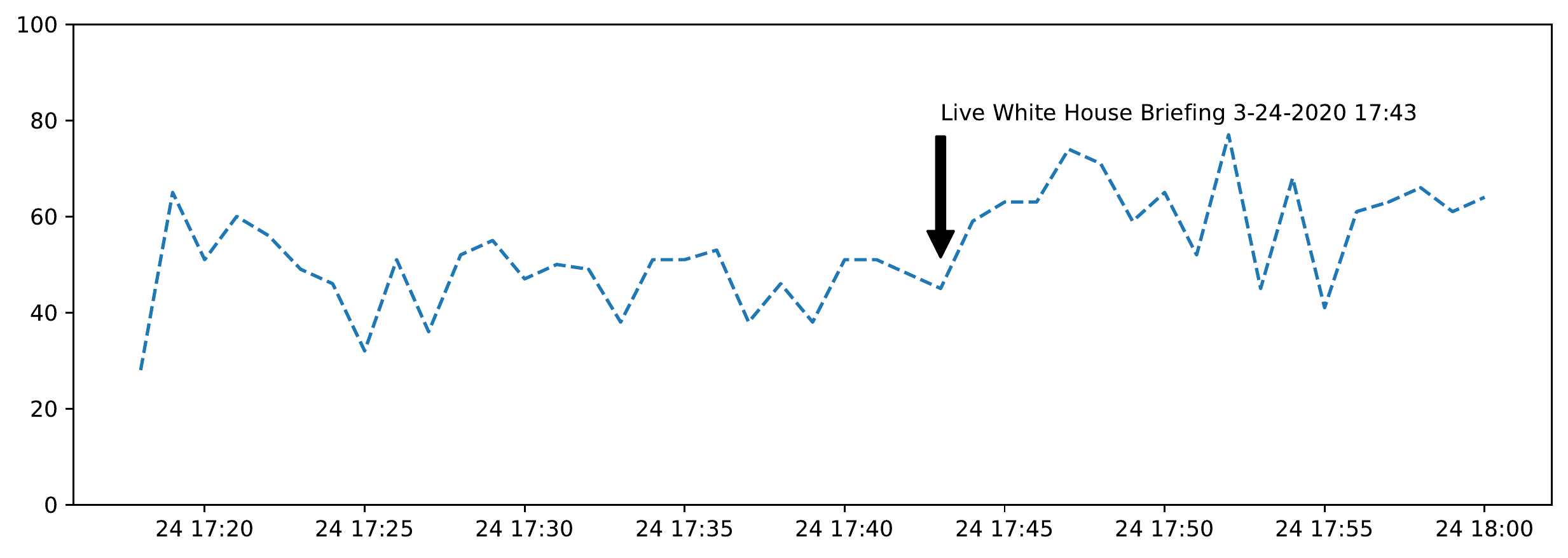}
  \label{a}
\end{subfigure}%
\newline
\begin{subfigure}{\textwidth}
\centering
  \includegraphics[width=0.75\linewidth]{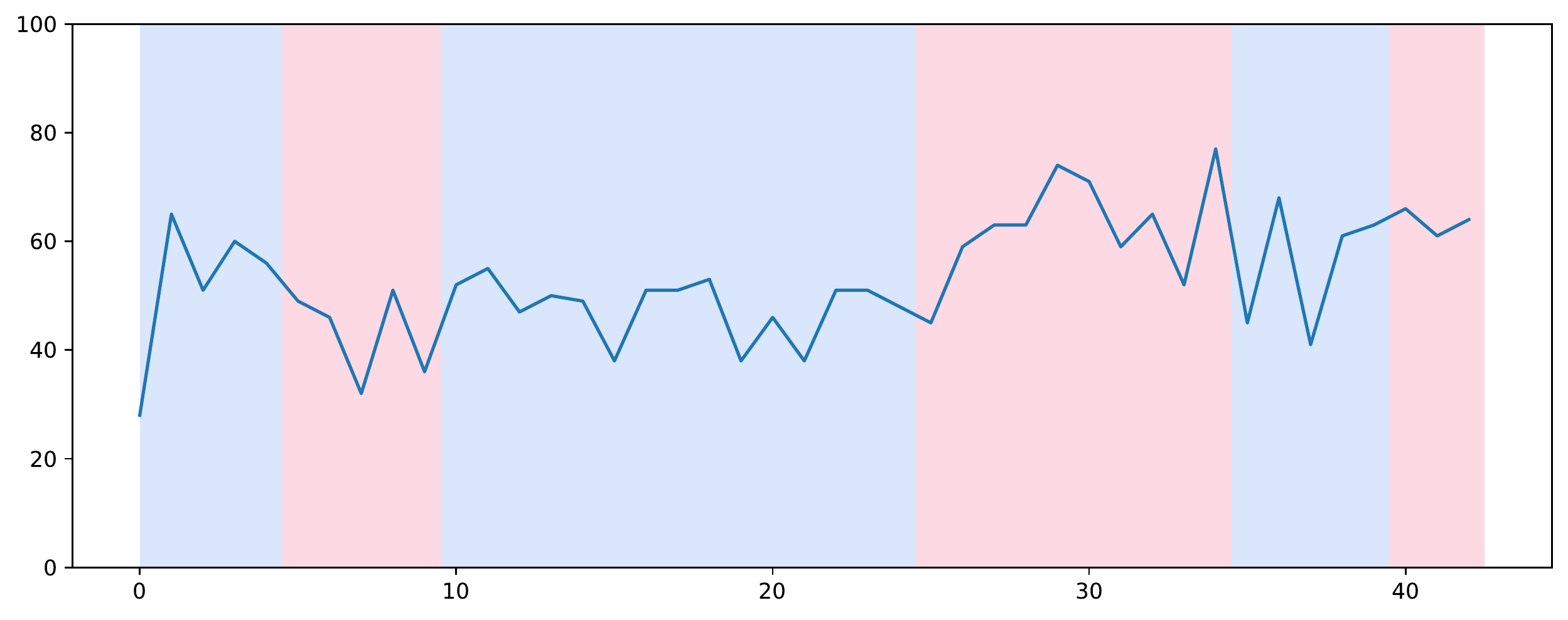}
  \label{b}
\end{subfigure}
\caption{Change Point Detection using Binary Segmentation for March 24, 2020}
\label{cpd_mar24}
\end{figure*}


When examining the distribution of the 20 topics across the corpora in \autoref{topics_dist}, Topics 18 ("potus"), 12 ("case.death.new"), 13 ("mask.ppe.ventil"), and 2 ("like.look.work") were the top five in the entire corpora.  For each plot, we labeled each topic with the first three terms of each topic for interpretability. In our trend analysis, we summed the number of tweets per minute, and then applied a moving weighted average of 60 minutes for topics March 24 - March 28, and 60 minutes for topics March 30 to April 8th. We provided two different plots in order to visualize smaller time frames such as March 24 of 44 minutes compared to longer time frames of 1477 for April 8. The results plotted in figures \autoref{baxes1} and \autoref{baxes2} show similar trends on a time-series basis per minute across the entire corpora of 5,506,223 tweets. These plots are in a style of "broken axes" \footnote{https://github.com/bendichter/brokenaxes} to indicate that the corpora are not continuous periods of time, but discrete time frames, which we selected to plot on one axis for convenience and legibility. We direct the reader to \autoref{datasets} for reference on the start and end datetimes, which are in UTC format, so please adjust accordingly for time zone. 

The x-axis denotes the number of minutes, where the entire corpora is 8463 total minutes of tweets. \autoref{baxes1} shows that for the corpora of March 24, 25, and 28, the topics (denoted in hash-marked lines) focused on Topic 18 "potus" and Topic 13 "mask.ppe.ventil" trended greatest. For the later time periods of March 30, March 31, April 4, 5 and 8 in \autoref{baxes2}, Topic 18 "potus"  and  Topic 13 "mask.ppe.ventil" (also in hash-marked lines) continued to trended high. It is also interesting that Topic 18 was never replaced as the top trending topic, across a span of 17 days (April 8, 2020 also includes early hours of April 9 2020 EST), potentially as this may have been a proxy for active government listening. The time series would temporally decrease in frequency during overnight hours, between the hours of midnight and 6:00 AM EST. But when examining the trend of the Topic 18 "potus" topic, we found that several live press briefings with the Coronavirus Task Force from @WhiteHouse would stimulate a spike in the Topic 18 topic 60 tweets per minute. 

\begin{itemize}
\item March 24, 2020, LIVE: Press Briefing with Coronavirus Task Force at 5:43 PM EST 
\item April 3, 2020, LIVE: Press Briefing with Coronavirus Task Force at 5:24 PM EST followed by a retweet from @WhiteHouse "Coronavirus—and we salute the great medical professionals on the front lines." at 5:59 PM EST
\item April 4, 2020, LIVE: Press Briefing with Coronavirus Task Force at 4:13 PM EST
\item April 5, 2020, LIVE: Press Briefing with Coronavirus Task Force at 6:53 PM EST
\item April 6, 2020, LIVE: Press Briefing with Coronavirus Task Force at 5:41 PM EST 
\item April 8, 2020: LIVE: Press Briefing with Coronavirus Task Force at 5:46 PM EST
\end{itemize}

We applied change point detection in the time series of tweets per minute for Topic 18 in the datasets March 24, 2020, April 3 - 4, 2020, April 5 - 6, 2020, and April 8, 2020, to identify whether the live press briefings coincided with inflections in time. Using the ruptures Python package \cite{truong2020selective} containing a variety of change point detection methods, we used binary segmentation \cite{killick2012optimal}, a standard method for change point detection. Given a sequence of data  ${y_{1:n} = (y_1, ..., y_n)}$ the model will have $m$ changepoints with their positions ${\tau_{1:m} = (\tau_1, ..., \tau_m)}$. Each changepoint position is an integer between 1 and $n-1$. The $m$ changepoints split the time series data into $m+1$ segments, with the \emph{i}th segment containing ${y_(\tau_{i-1}+1):\tau_i}$. Changepoints are identified by minimizing a cost function, $C$ for a given segment, where ${\beta f(m)}$ is a penalty to prevent overfitting. 

\[\sum_{i=1}^{m+1}[C(y_(\tau_{i-1}+1):\tau_i)] + \beta f(m) \] where twice the negative log-likelihood is a commonly used cost function.

Binary segmentation detects multiple changepoints across the time series by repeatedly testing on different subsets of the sequence. It checks to see if a $\tau$ exists that satisfies: 
\[C(y_{1:\tau} + C(y_{(\tau+1):n}) + \beta < C(y_{1:n}) \]

If not, then no changepoint is detected and the method stops. But if a changepoint is detected, the data are split into two segments consisting of the time series before (\autoref{cpd_mar24} blue) and after (\autoref{cpd_mar24} pink) the changepoint. We can clearly see in \autoref{cpd_mar24} that the timing of the White House briefing indicates a changepoint in time, giving us the intuition that this briefing influenced an uptick in the the number of tweets. We provide additional examples in the Appendix.

Our topic findings are consistent with the published analyses on Covid19 and Twitter, such as \cite{singh2020first} who found major themes of healthcare and illness and international dialogue, as we noticed in our four non-English topics. They are also similar to by Thelwall et al. \cite{thelwall2020retweeting} who manually reviewed tweets from a corpus of 12 million tweets occurring earlier and overlapping our dataset (March 10 - 29). Similar topics from their findings to ours includes "lockdown life", "politics", "safety messages", "people with COVID-19", "support for key workers", "work", and "COVID-19 facts/news". 

Further, our dataset of Covid19 tweets from March 24 to April 8, 2020 occurred during a month of exponential case growth.  By the end of our data collection period, the number of cases had increased by 7 times to 427,460 cases on April 8, 2020 \cite{CDC_cases}. The key topics we identified using our multiple methods were representative of the public conversations being had in news outlets during March and April, including: 
\begin{itemize}
\item CDC allowing private companies to make tests (March 3, 2020)
\item President Trump declaring Covid19 a national emergency (March 13, 2020)
\item CDC advising against social gatherings of more than 50 people (March 15, 2020)\cite{CDC_Mass}
\item CDC issuing (March 17, 2020) Strategies for Optimizing the Supply of Facemasks
\item President Trump mentioning hydroxychloroquine as a potential Covid19 treatment.\cite{liptak_2020}
\item The HHS Assistant Secretary for Health and U.S. Surgeon General issuing a letter to the healthcare community to optimize ventilator use (March 31, 2020)
\item The White House issues a Memorandum on Order Under the Defense Production Act Regarding the Purchase of Ventilators (April 2, 2020)\cite{whitehouse_memo}
\item CDC issuing guidance on wearing facial coverings (April 3, 2020) \cite{CDC_masks}
\end{itemize}

\section{Uniform Manifold Approximation and Projection}
Term-frequency inverse-document-frequency (TF-IDF)\cite{ramos2003using} is a weight that signifies how valuable a term is within a document in a corpus, and can be calculated at the n-gram level. TF-IDF has been widely applied for feature extraction on tweets used for text classification \cite{lee2011twitter} \cite{hong2011predicting}, analyzing sentiment \cite{barnaghi2016opinion}, and for text matching in political rumor detection \cite{jin2017detection} With TF-IDF, unique words carry greater information and value than common, high frequency words across the corpus. TF-IDF can be calculated as follows:

\[w_{i,j}  = tf_{i,j} \times \log_\frac{N}{df_i}\]

Where $i$ is the term, $j$ is the document, and $N$ is the total number of documents in the corpus. TF-IDF calculates the term frequency $tf_{i,j}$ multiplied by the log of the inverse document frequency $\frac{N}{df_i}$. The term frequency $tf_{i,j}$ is calculated as the frequency of $i$ in $j$ divided by all terms $i$ in given $j$. The inverse document frequency is $\frac{N}{df_i}$ is the log of the total number of documents $j$ in the corpus divided by the number of documents $j$ containing term, $i$. 

Using the Scikit-Learn implementation of TfidfVectorizer and setting max\_features to 10000, we transformed our corpus of 5,506,223 tweets into a $\mathbb{R}^{n \times k}$ sparse dimensional matrix of shape (5506223, 10000).   Note, prior to fitting the vectorizer, our corpus of tweets was pre-processed during the keyword analysis stage. We chose to visualize how the 20 topics grouped together using Uniform Manifold Approximation and Projection (UMAP) \cite{mcinnes2018umap}.  UMAP is a dimension reduction algorithm that finds a low dimensional representation of data with similar topological properties as the high dimensional space. It measures the local distance of points across a neighborhood graph of the high dimensional data, capturing what is called a fuzzy topological representation of the data. Optimization is then used to find the closest fuzzy topological structure by first approximating nearest neighbors using the Nearest-Neighbor-Descent algorithm and then minimizing local distances of the approximate topology using stochastic gradient descent \cite{mcinnes}. When compared to t-Distributed Stochastic Neighbor Embedding (t-SNE), UMAP has been observed to be faster \cite{coenen_pearce} with clearer separation of groups.  

Due to compute limitations in fitting the entire high dimensional vector of nearly 5.5M records, we randomly sampled one million records. We created an embedding of the vectors along two components to fit the UMAP model with the Hellinger metric which compares distances between probability distributions, as follows:

\[h(P, Q) = \frac{1}{\sqrt{2}} \cdot \left\Vert\left(\sqrt{P} - \sqrt{Q}\right)\right\Vert_2\]
We visualized the word vectors with their respective labels, which were the assigned topics generated from the LDA model.  We used the default parameters of n\_neighbors = 15 and min\_dist = 0.1. \autoref{umap} presents the visualization of the TF-IDF word vectors for each of the 1 million tweets with their labeled topics. UMAP is supposed to preserve local and global structure of data, unlike t-SNE that separates groups but does not preserve global structure. As a result, UMAP visualizations intend to allow the reader to interpret distances between groups as meaningful. In \autoref{umap} each topic is color-coded by its respective topic. 

The UMAP plots appear to provide further evidence of the quality and number of topics generated. Our observations is that many of these topic "clusters" appear to have a single dominant color indicating distinct grouping. There is strong local clustering for topics that were also prominent in the keyword analysis and topic modeling time series plots. A very distinct and separated mass of purple tweets represents the "100: N/A" topic which is an undefined topic. This means that the LDA model outputted equal scores across all 20 topics for any single tweet. As a result, we could not assign a topic to these tweets because they all had uniform scores.  But this visualization informs us that the contents of these tweets were uniquely distinct from the others. Examples of tweets in this "100: N/A" cateogry include  "See, \#Democrats are always guilty of whatever", "Why are people still getting in cruise ships?!?", "Thank you Mike you are always helping others and sponsoring Anchors media shows.", "We cannot let this woman’s brave and courageous actions go to waste! \#ChinaLiedPeopleDied \#Chinaneedstopay", "I wish people in this country would just stay the hell home instead of GOING TO THE BEACH". Other observations reveal that the mask-related topic 10 in purple, and potentially a combination of 8 and 9 in red are distinct from the mass of noisy topics in the center of the plot. We can also see distinct separation of aqua-colored topic 18 "potus" and potentially topics 5 and 6 in yellow.

We refer the reader to other examples where UMAP has been leveraged for Twitter analysis, to include Darwish et al. \cite{darwish2019unsupervised} for identifying clusters of Twitter users with controversial topic similarity, Vargas \cite{vargas2019event} for event detection, political polarization by Darwish et al. \cite{darwish2019unsupervised} and estimating political leaning of users by \cite{stefanov2019predicting}.  


\begin{figure*}
\centering
\begin{subfigure}{.3\textwidth}
  \centering
  \includegraphics[width=.8\linewidth]{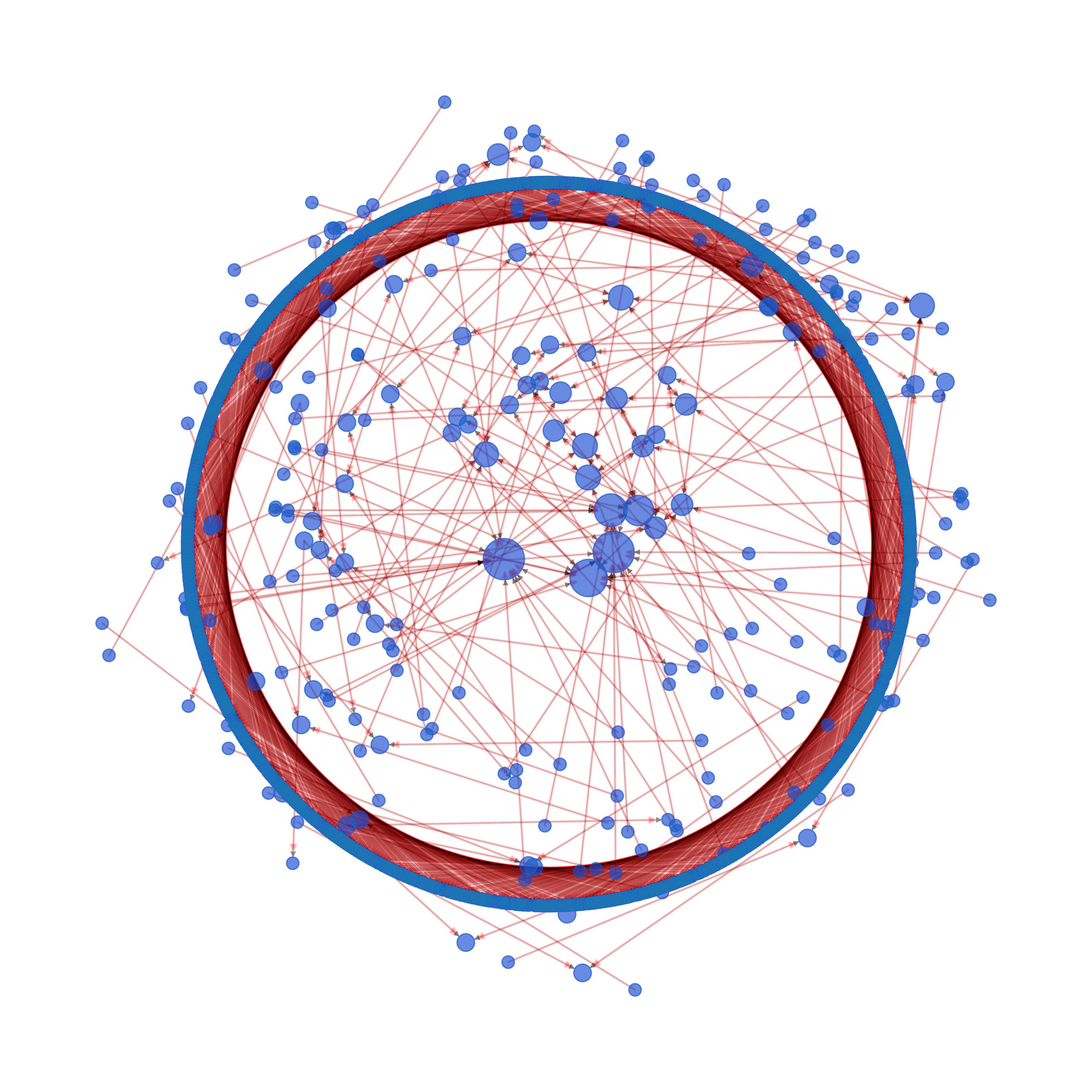}
  \caption{G1 at 19 seconds}
  \label{8a}
\end{subfigure}%
\begin{subfigure}{.3\textwidth}
  \centering
  \includegraphics[width=.8\linewidth]{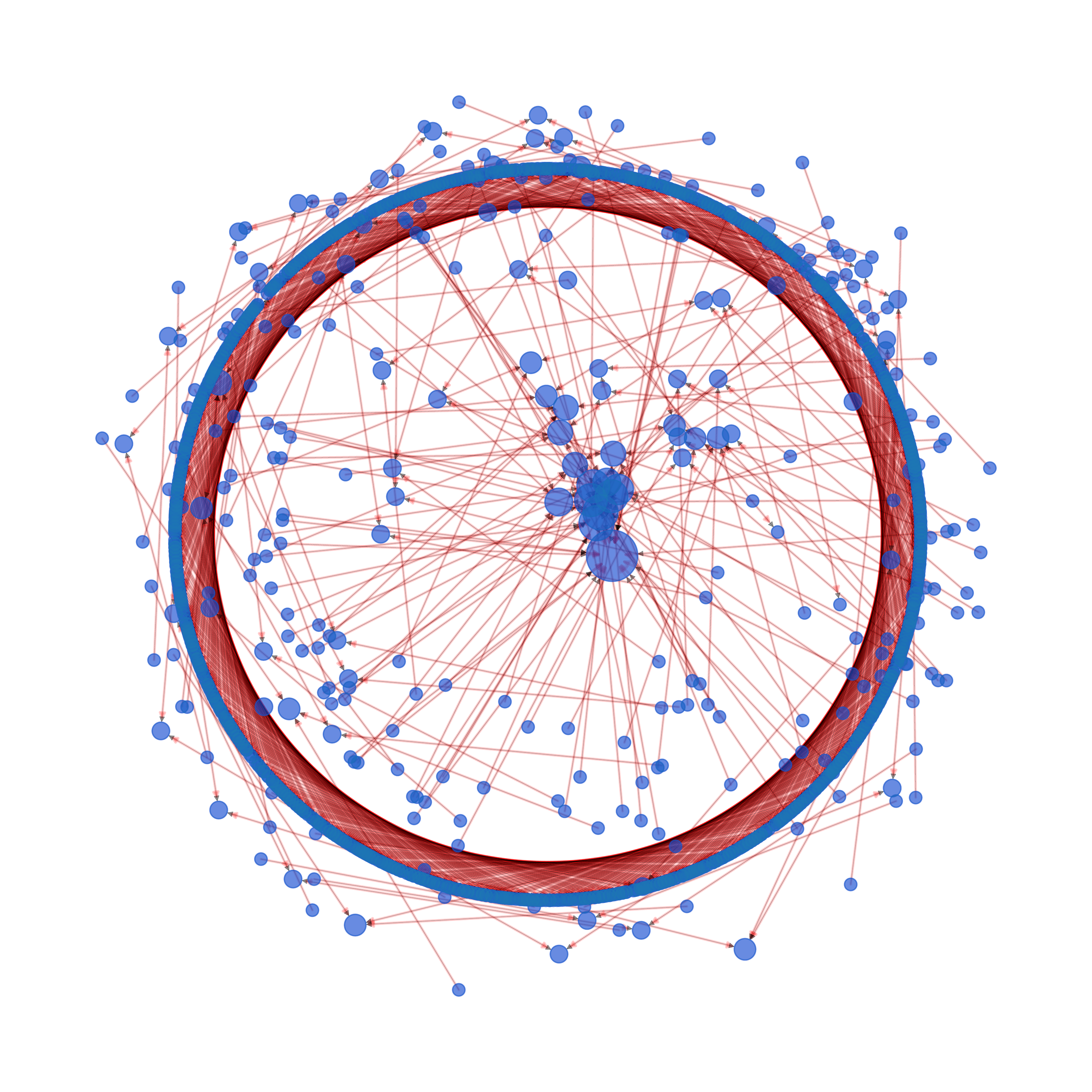}
  \caption{G2 at 5.47 min.}
  \label{b}
\end{subfigure}
\begin{subfigure}{.3\textwidth}
  \centering
  \includegraphics[width=.8\linewidth]{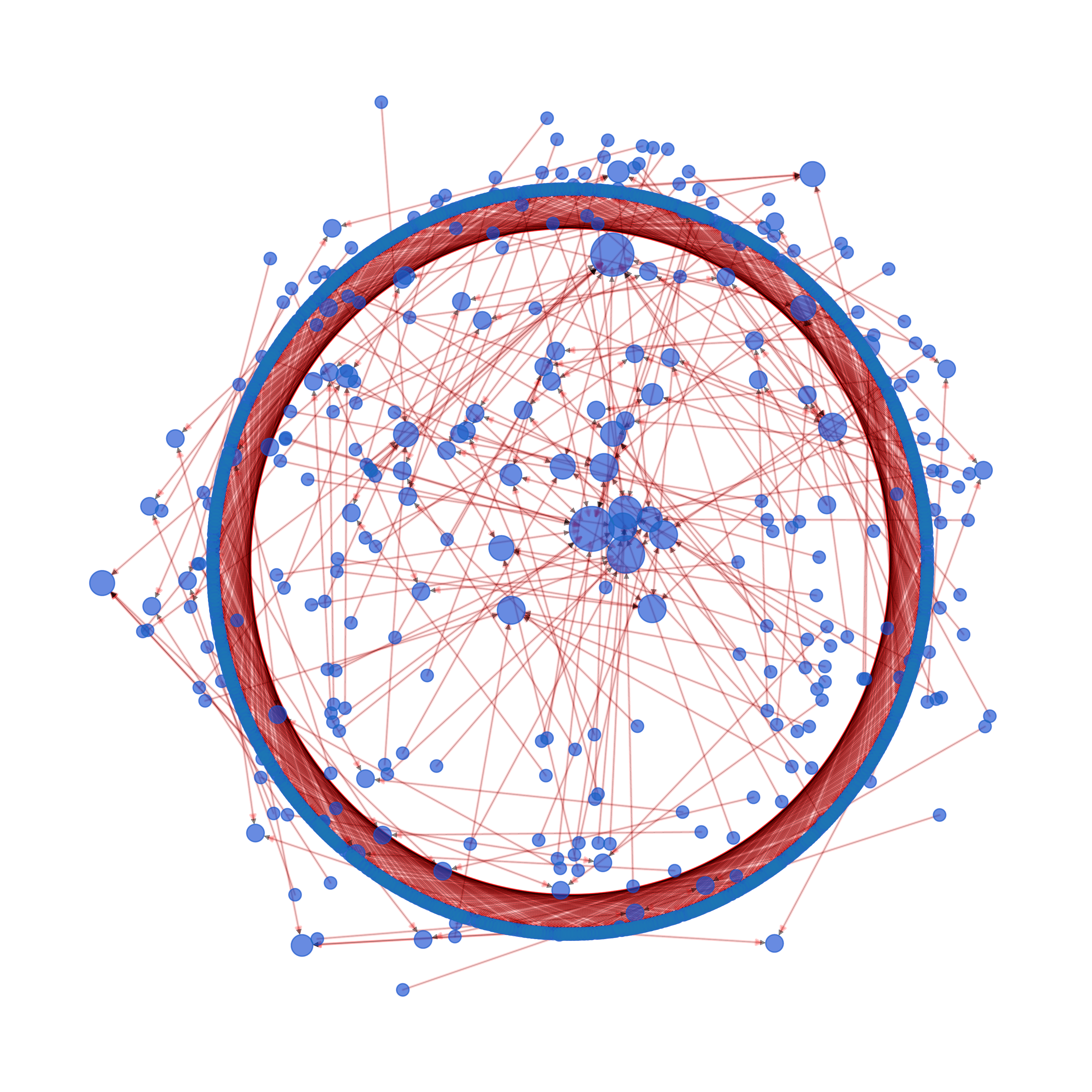}
  \caption{G3 at 9.85 min.}
  \label{c}
\end{subfigure}

\begin{subfigure}{.3\textwidth}
  \centering
  \includegraphics[width=.8\linewidth]{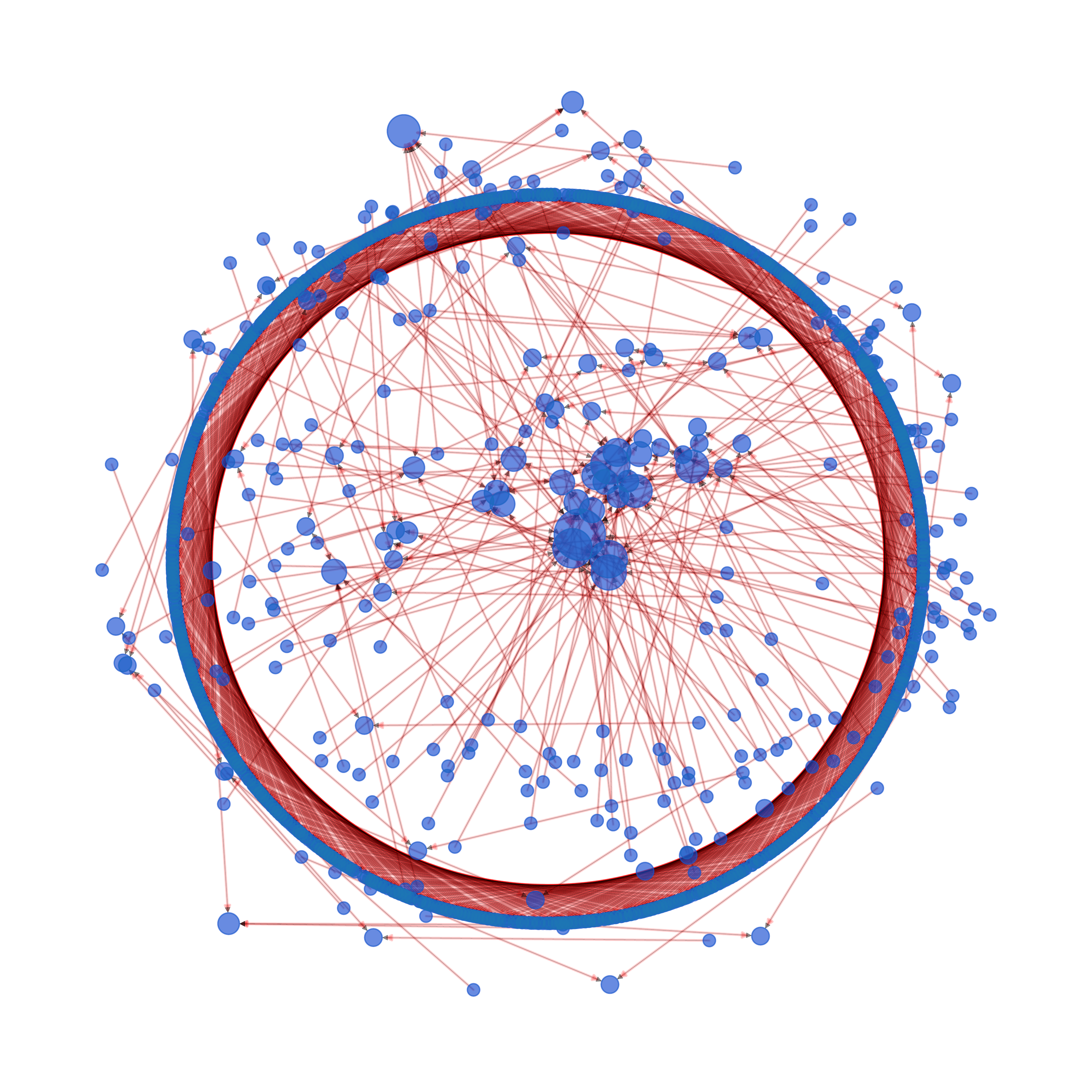}
  \caption{G4 at 14.76 min.}
  \label{d}
\end{subfigure}
\begin{subfigure}{.3\textwidth}
  \centering
  \includegraphics[width=.8\linewidth]{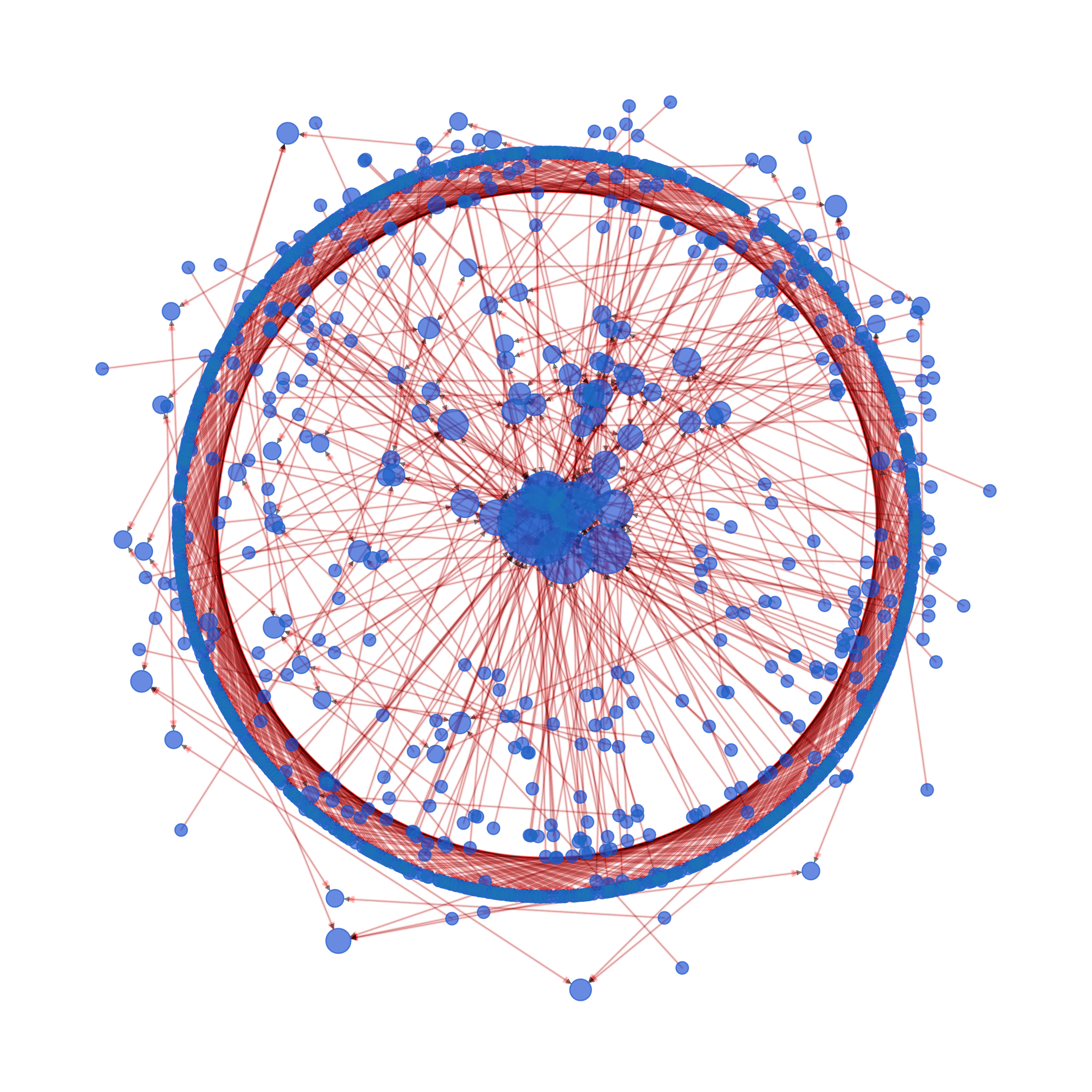}
  \caption{G5 at 60 min.}
  \label{e}
\end{subfigure}
\begin{subfigure}{.3\textwidth}
  \centering
  \includegraphics[width=.8\linewidth]{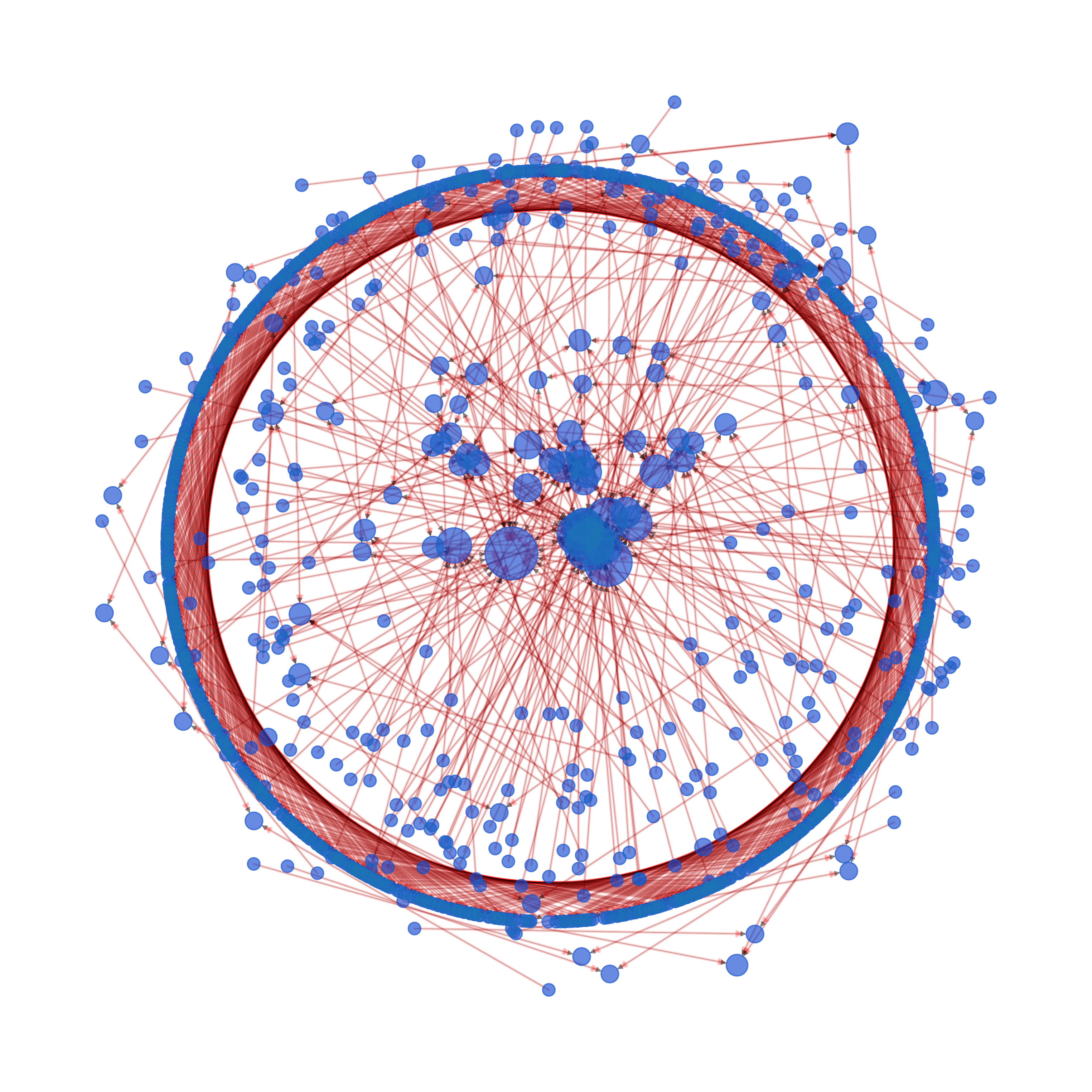}
  \caption{G6 at 2.78 hr.}
  \label{f}
\end{subfigure}

\begin{subfigure}{.3\textwidth}
  \centering
  \includegraphics[width=.8\linewidth]{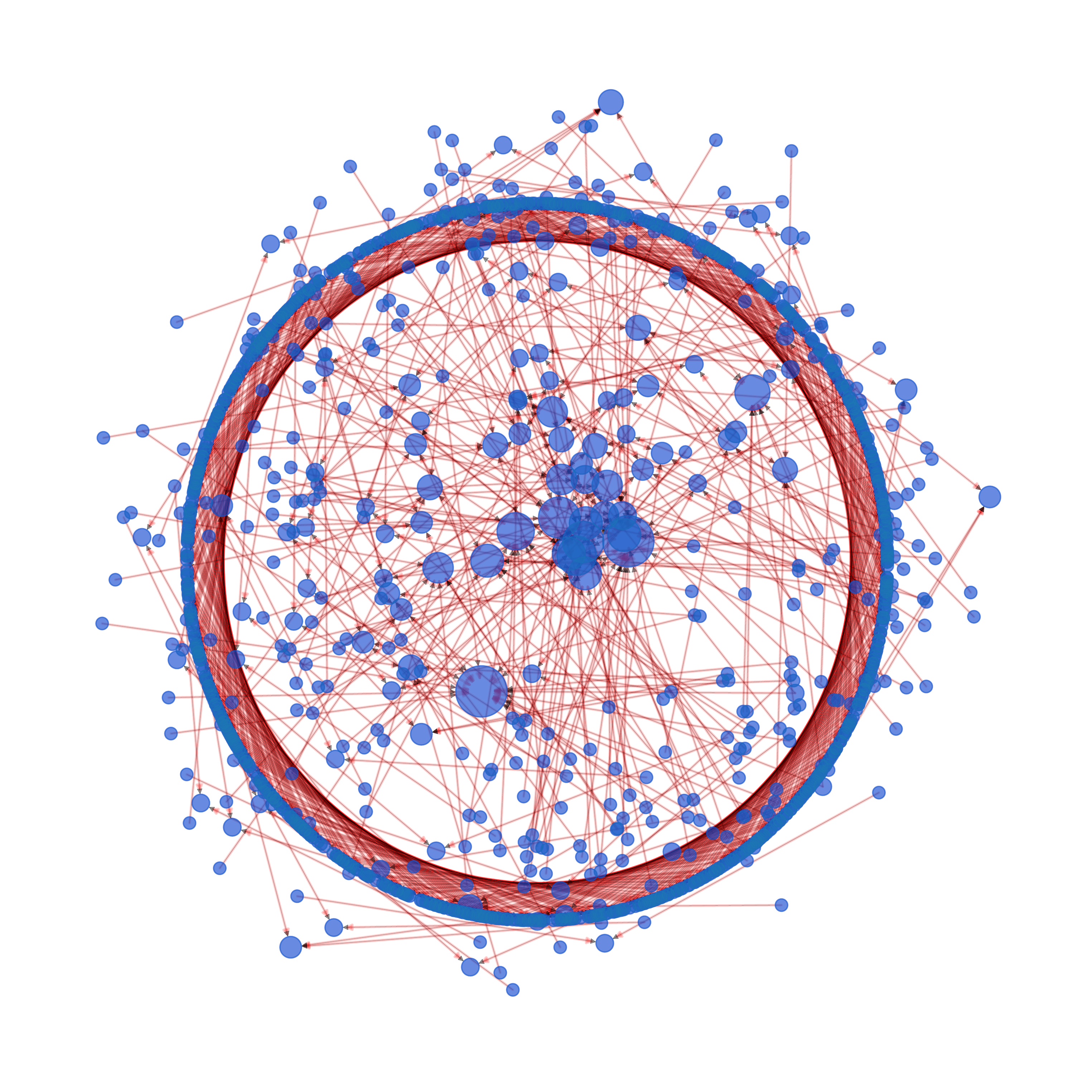}
  \caption{G7 at 3.7 hr.}
  \label{g}
\end{subfigure}
\begin{subfigure}{.3\textwidth}
  \centering
  \includegraphics[width=.8\linewidth]{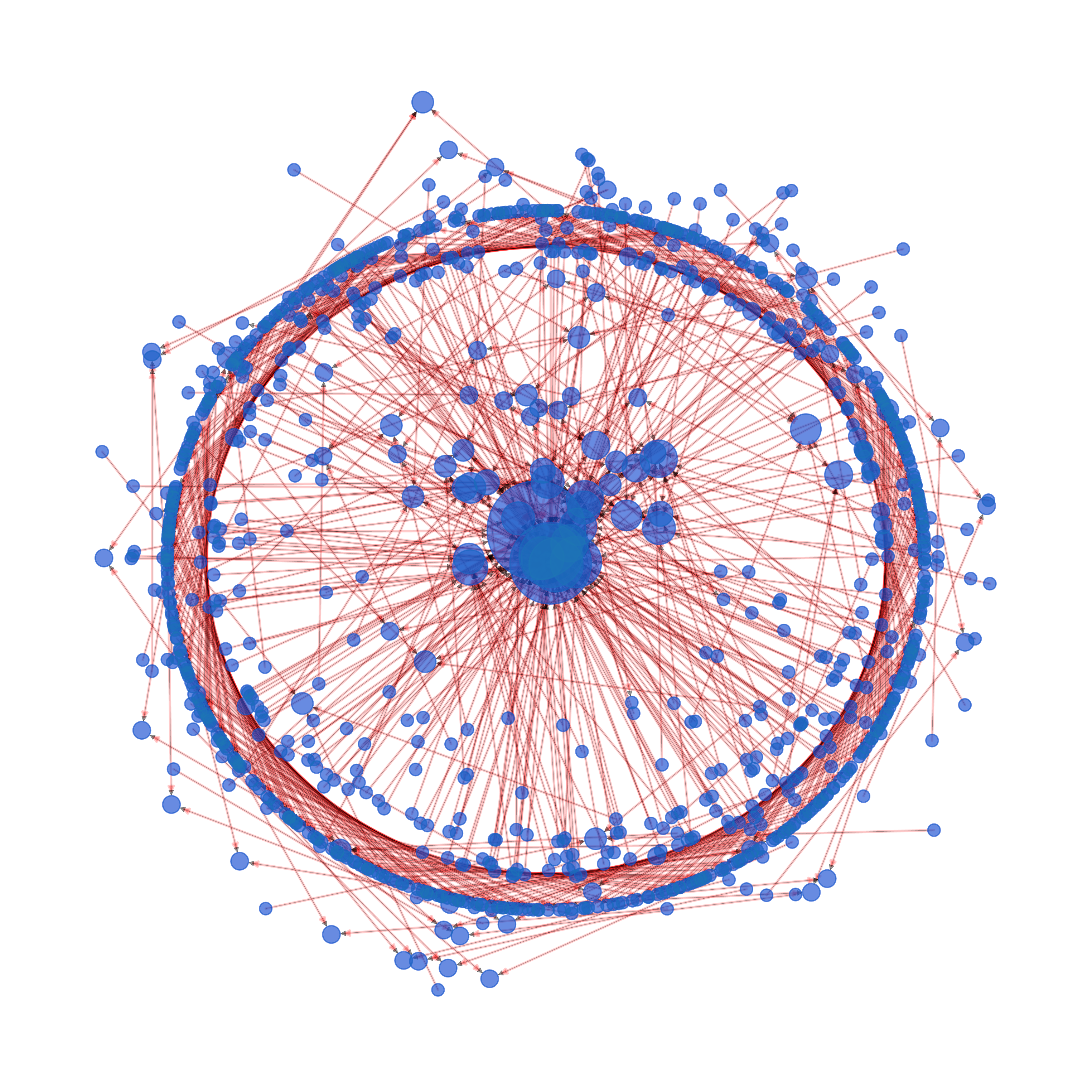}
  \caption{G8 at 24 hr.}
  \label{h}
\end{subfigure}
\begin{subfigure}{.3\textwidth}
  \centering
  \includegraphics[width=.8\linewidth]{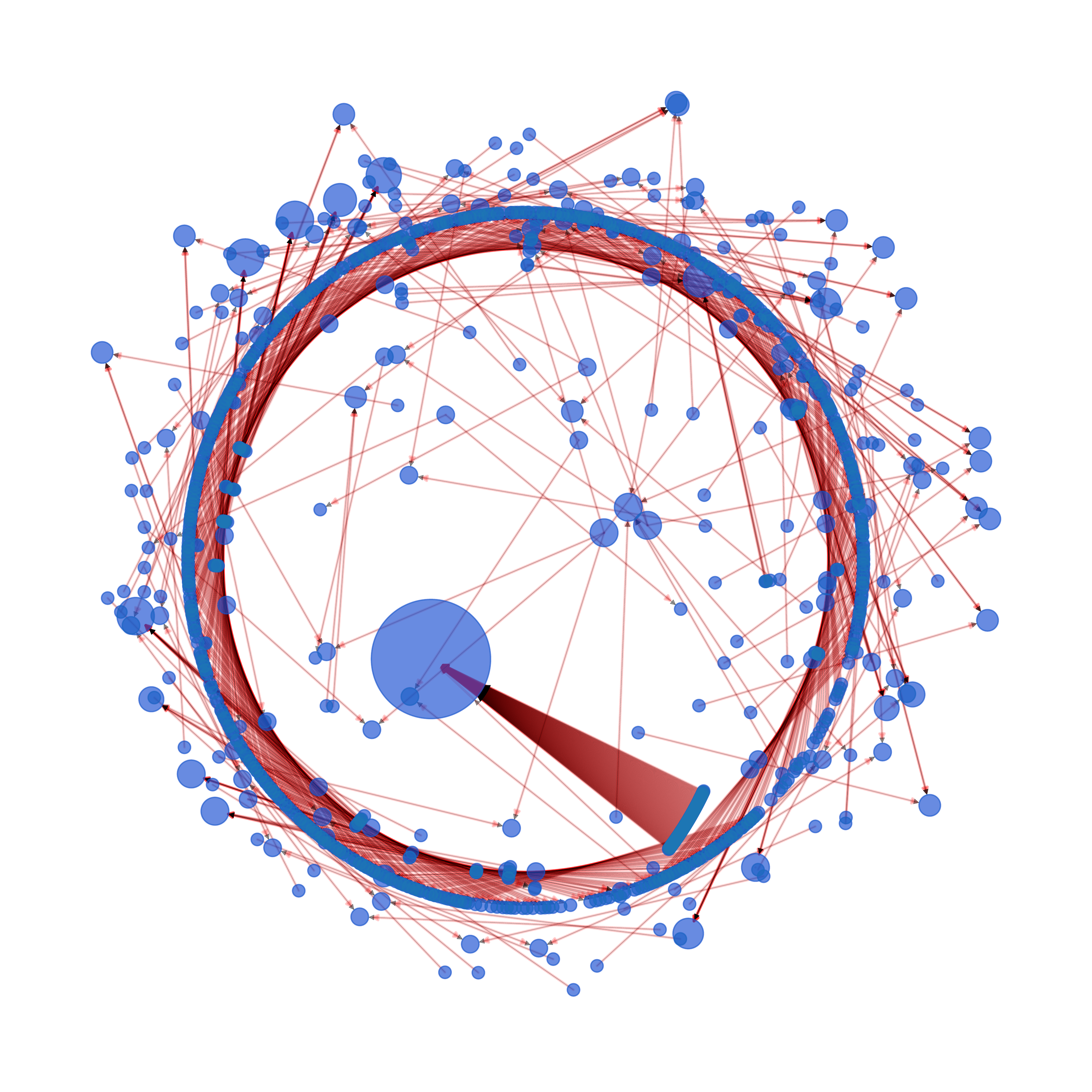}
  \caption{G9 at 168 hr.}
  \label{i}
\end{subfigure}

\caption{Directed Graphs of Covid19 Retweeting Activity at Nine Different Points in Time (G1 - G9) between March 24 - March 28th using the Kamada Kawai Layout}
\label{networkGraphs}
\end{figure*}


\section{Time-to-Retweet Analysis}

\begin{figure*}[t]
\centering
\includegraphics[width=1.0\textwidth]{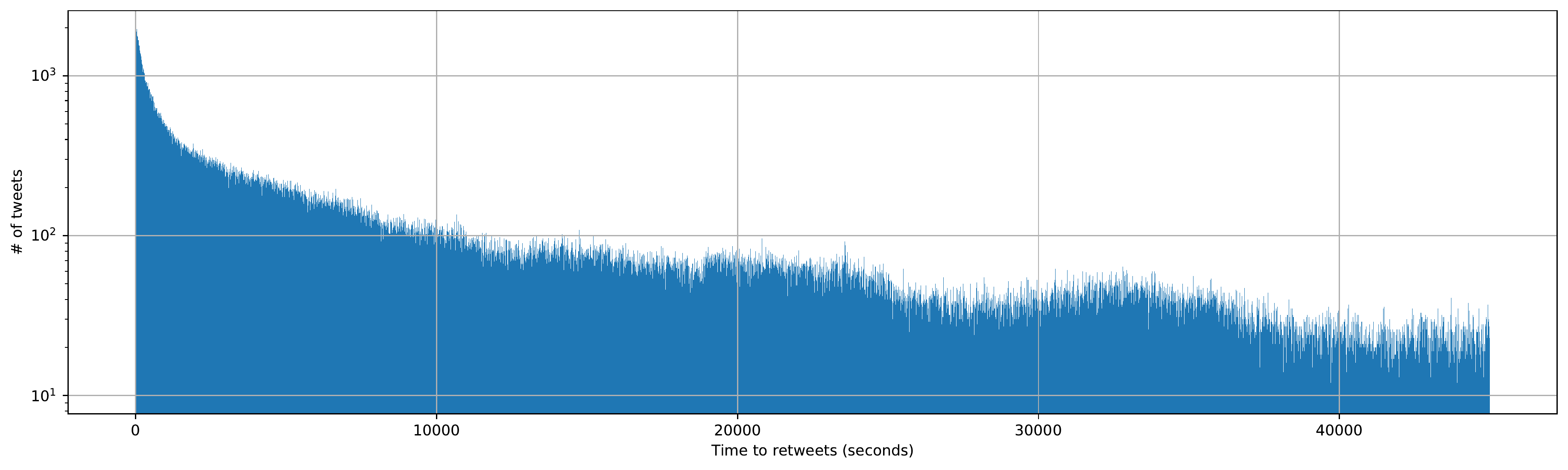}
\caption{Seconds to Retweet, March 24 - 28th Corpora }
\label{secondsrt}
\end{figure*}

\begin{figure*}[t]
\centering
\includegraphics[width=1.0\textwidth]{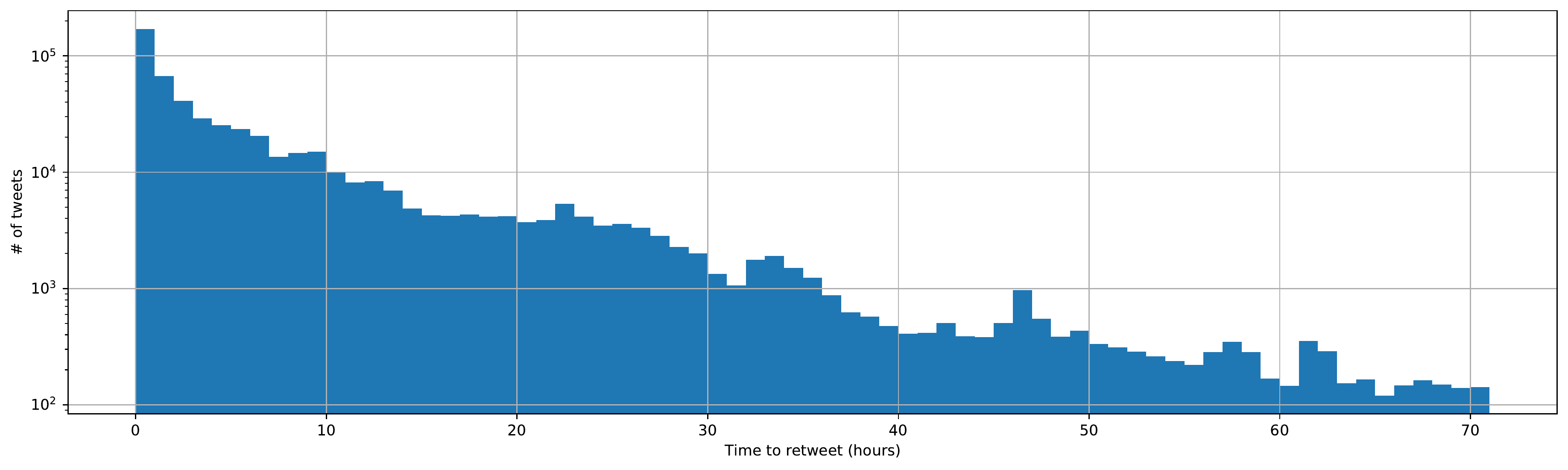}
\caption{Hours to Retweet, March 24 - 28th Corpora }
\label{hoursrt}
\end{figure*}

Retweeting is a special activity reserved for Twitter where any user can "retweet" messages which allows them to disseminate their messages rapidly to their followers.  Further, a highly retweeted tweet might signal that an issue has attracted attention in the highly competitive Twitter environment, and may give insight about issues that resonate with the public \cite{naveed2011bad}. Whereas in the first three analyses we used no retweets, in the time-series and network modeling that follows, we exclusively use retweets. We began by measuring time-to-retweet. Wang et al. \cite{wang2017crisis} calls this "response time" and used it to measure response efficiency and speed of information dissemination during Hurricane Sandy. Wang analyzed 986,579 tweets and found that 67\% of re-tweets occur within 1 h \cite{wang2017crisis}. We researched how fast other users retweet in emergency situations, such as what Spiro \cite{spiro2012waiting} reported for natural disasters, and how Earle \cite{earle2010omg} reported as 19 seconds for retweeting about an earthquake. 

We extracted metadata from our corpora for the Tweet, User, and Entities objects. For reference, we direct the reader to the Twitter Developer guide that provides a detailed overview of each object\cite{twitter}.  Due to compute limitations, we selected a sample that consisted of 736,561 tweets that included retweets from the corpora of March 24 - 28, 2020.  However, since we were only focused on retweets, out of the corpus of 736,561 tweets, we reduced it to 567,909 (77\%) that were only retweets. The metadata we used for both our Time-to-Retweet and Directed Graph analyses in the next section, included:

\begin{enumerate}
\item Created\_at  (string) - UTC time when this Tweet was created.
\item Text (string) - The actual UTF-8 text of the status update. See twitter-text for details on what characters are currently considered valid. 
\item From the User object, the id\_str (string) - The string representation of the unique identifier for this User. 
\item From the retweeted\_status object (Tweet) - the created\_at UTC time when the Retweet was created.
\item From the retweeted\_status object (Tweet) - the id\_str which is the unique identifier for the retweeting user.
\end{enumerate}

We used the corpus of retweets and analyzed the time between the tweet created\_at and the retweeted created\_at.  

\[time\_to\_rt = rt\_object - tw\_object\] 

Here, the rt\_object is the datetime in UTC format for when the message that was retweeted was originally posted. The tw\_object is the datetime in UTC format when the current tweet was posted. As a result, the datetime for the rt\_object is older than the datetime for the current tweet. This measures the time it took for the author of the current tweet to retweet the originating message. This is similar to Kuang et al. \cite{kuang2014predicting} who defined response time of the retweet to be the time difference between the time of the first retweet and that of the origin tweet. Further, Spiro et al. \cite{spiro2012waiting} calls these "waiting times". The median time-to-retweet for our corpus was 2.87 hours meaning that half of the tweets occurred within this time (less than what Wang reported as 1.0 hour), and the mean was 12.3 hours. \autoref{secondsrt} shows the histogram of the number of tweets by their time to retweet in seconds and \autoref{hoursrt} shows it in hours. 

\begin{figure}[ht]
\centering
\includegraphics[width=\columnwidth]{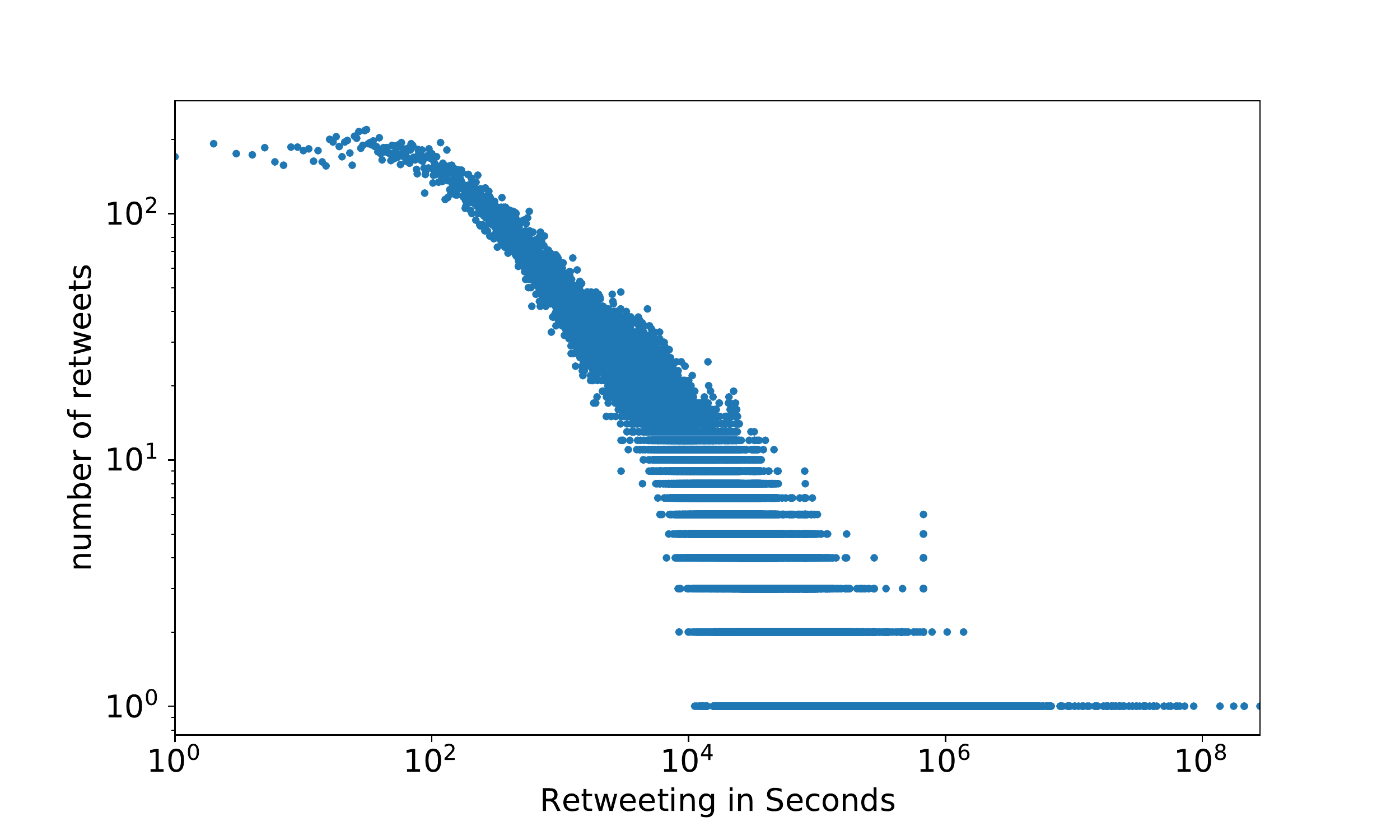}
\caption{Log Distribution of Covid19 Retweets from March 24 - 28, 2020}
\label{rt_logtweets}
\end{figure}

\begin{figure}[ht]
\centering
\includegraphics[width=\columnwidth]{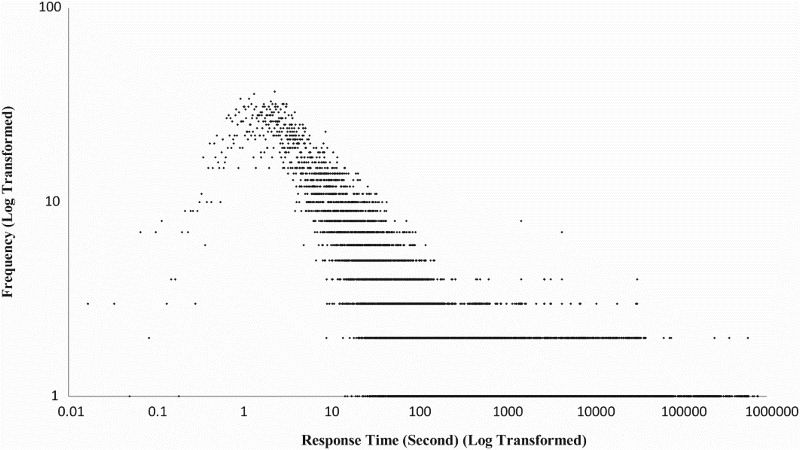}
\caption{Log Distribution of H7N9-related messages on Sina Weibo, March 2013}
\label{zhang_rt}
\end{figure}

Further, we found that compared to the 2013 Avian Influenza outbreak (H7N9) in China described by Zhang et al. \cite{zhang2017social} Covid19 retweeters sent more messages earlier than H7N9. Zhang analyzed the log distribution of 61,024 H7N9-related posts during April 2013 and plotted reposting time of  messages on Sina Weibo, a Chinese Twitter-like platform and one of the largest microblogging sites in China \autoref{zhang_rt}. Zhang found that H7N9 reposting occurred with a median time of 222 minutes (i.e. 3.7 hours) and a mean of 8520 minutes (i.e. 142 hours). Compared to Zhang's study, we found our median retweet time to be 2.87 hours, about 50 minutes faster than the reposting time during H7N9 of 3.7 hours. When comparing \autoref{rt_logtweets} and \autoref{zhang_rt}, it appears that Covid19 retweeting does now completely slow down until 2.78 hours later ($10^4$ seconds). For H7N9 it appears to slow down much earlier by 10 seconds.

Unfortunately few studies appear to document retweeting times during infectious disease outbreaks which made it hard to compare how Covid19 retweeting behavior against similar situations. Further, the H7N9 outbreak in China occurred seven years ago and may not be a comparable set of data for numerous reasons. Chinese social media may not represent similar behaviors with American Twitter and this analysis does not take into account multiple factors that imply retweeting behavior to include the context, the user's position, and the time the tweet was posted \cite{naveed2011bad}.

\subsection{TF-IDF Message and User Description Features of Rapid Retweeters}
\begin{figure*}[ht]
\centering
\begin{minipage}[b]{.4\textwidth}
\includegraphics[width=0.9\textwidth]{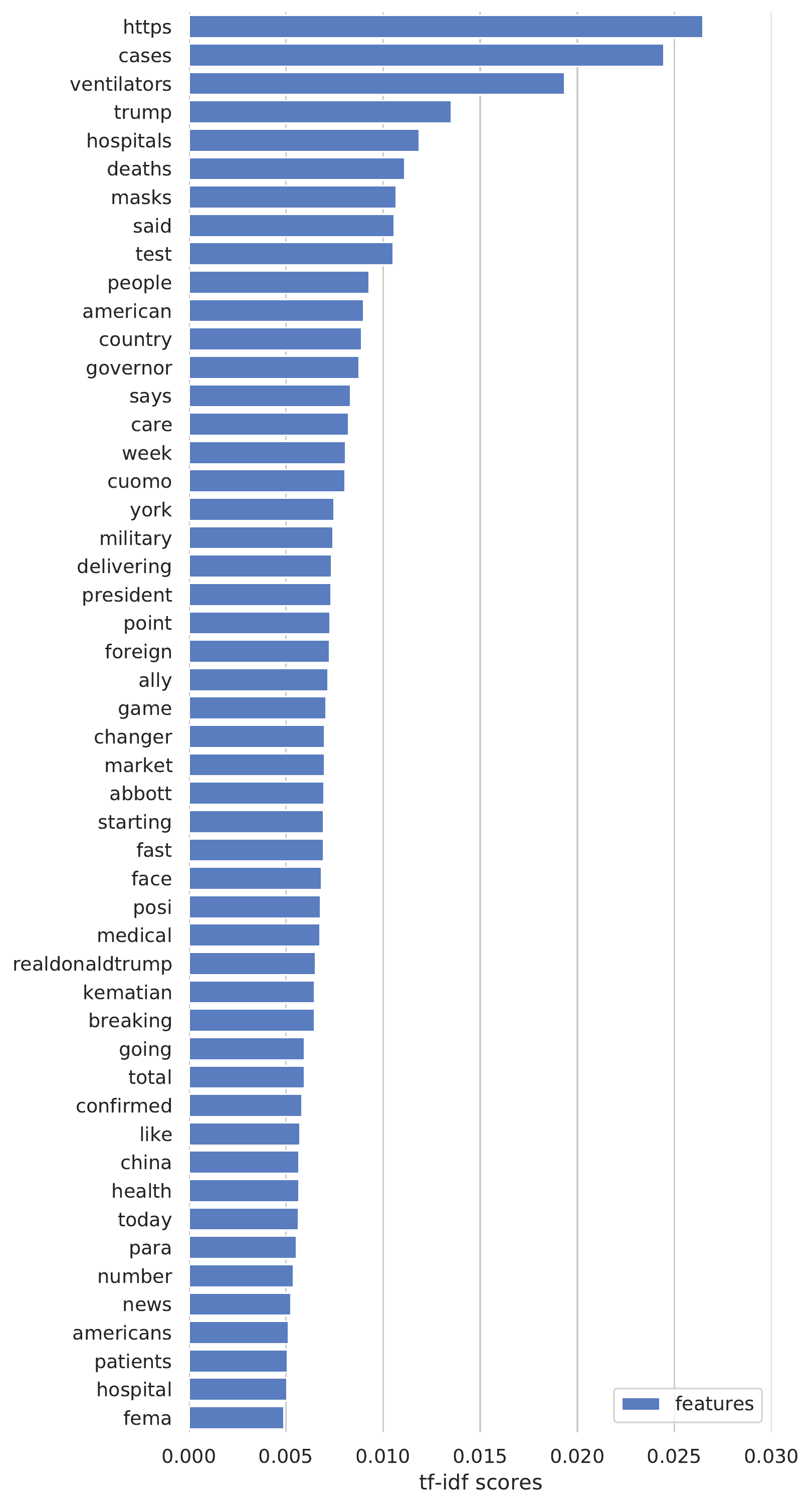}
\caption{TF-IDF Scores for Rapidly Retweeted Messages}
\end{minipage}\qquad
\begin{minipage}[b]{.4\textwidth}
\includegraphics[width=0.9\textwidth]{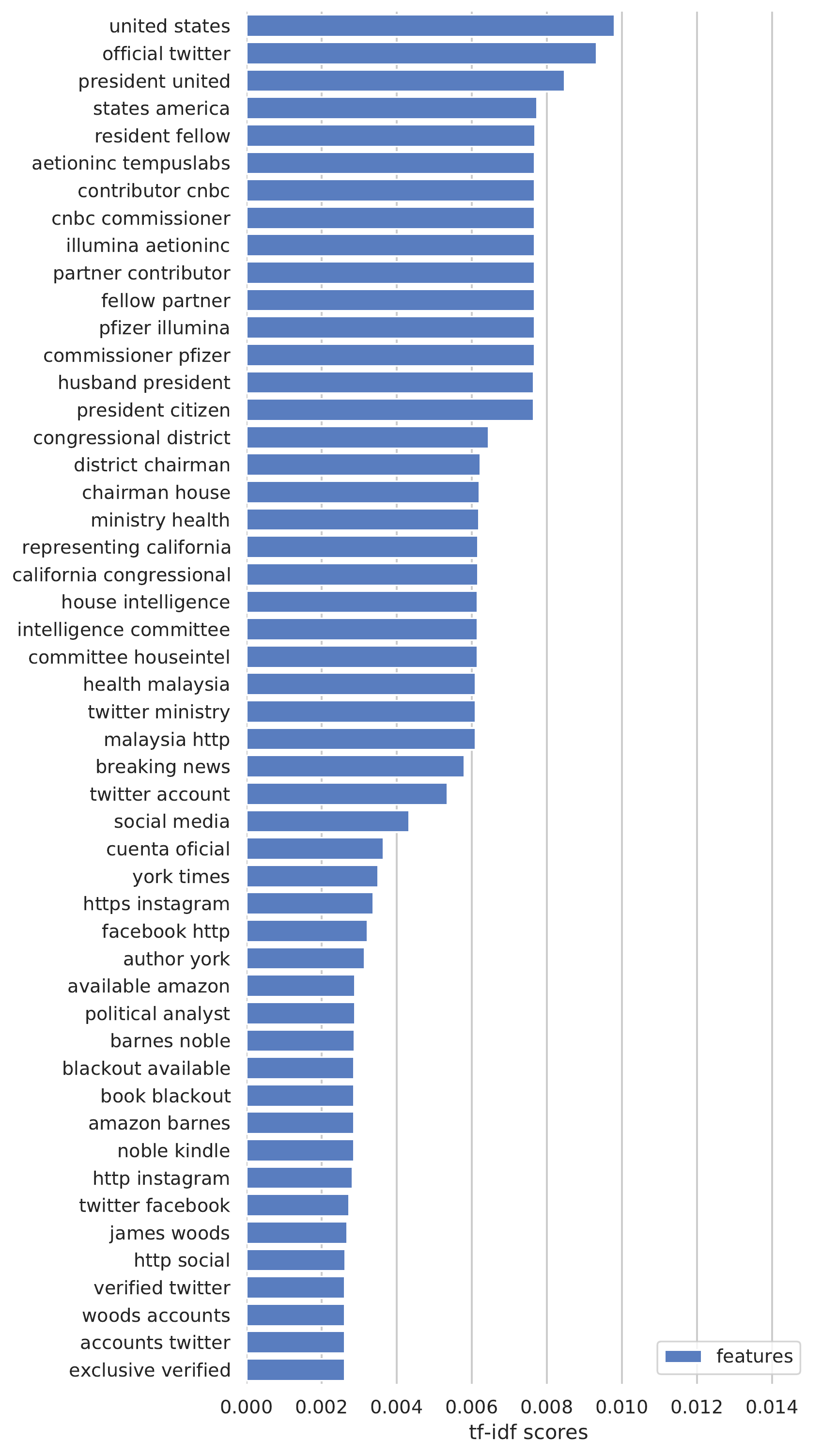}
\caption{TF-IDF Scores for Descriptions of Retweeted Users}
\label{tfidf_appendix}
\end{minipage}

\end{figure*}

We also analyzed what rapid retweeters, or those retweeting messages even faster than the median, in less than 10,000 seconds were saying.  In \autoref{tfidf_appendix} we plotted the top 50 TF-IDF features by their scores for the text of the retweets.  It is intuitive to see that URLs are being retweeted quickly by the presence of "https" in the body of the retweeted text.  This is also consistent with studies by Suh et al. \cite{suh2010want} who indicated that tweets with URLs were a significant factor impacting retweetability. We found terms that were frequently mentioned during the early-stage keyword analysis and topic modeling mentioned again: "cases", "ventilators", "hospitals", "deaths", "masks", "test", "american", "cuomo", "york", "president", "china", and "news".  When analyzing the descriptions of the users who were retweeted in \autoref{tfidf_appendix}, we ran the TF-IDF vectorizer on bigrams in order to elicit more interpretable terms. User accounts whose tweets were  rapidly retweeted, appeared to describe themselves as political, news-related, or some form of social media account, all of which are difficult to verify as real or fake. 

\section{Network Modeling}

\begin{table}[]
\centering
\caption{Statistics about Each Network Community}
\label{graph_stats_table}
\resizebox{\columnwidth}{!}{%
\begin{tabular}{@{}llllllll@{}}
\toprule
Graphs & Ranking Speed & Time Point & Density & Nodes & 1st & 2nd & 3rd \\ \midrule
G1 & 1 & 19 sec & 0.000428 & 1278 & 11 & 11 & 9 \\
G2 & 2 & 328 sec (5.47 min) & 0.000449 & 1248 & 17 & 8 & 8 \\
G3 & 3 & 591 sec (9.85 min) & 0.000450 & 1247 & 13 & 12 & 9 \\
G4 & 4 & 885.6 sec (14.76 min) & 0.000460 & 1234 & 17 & 10 & 10 \\
G5 & 5 & 3600 sec (60 min) & 0.000567 & 1110 & 41 & 27 & 20 \\
G6 & 6 & 10000 sec (2.78 hrs) & 0.000538 & 1139 & 18 & 15 & 15 \\
G7 & 7 & 13,320 sec (3.7 hrs) & 0.000540 & 1138 & 17 & 17 & 11 \\
G8 & 8 & 86,400 sec (24 hrs) & 0.000685 & 1005 & 63 & 43 & 26 \\
G9 & 9 & 604,800 sec (1 week) & 0.000598 & 1067 & 92 & 9 & 9 \\ \bottomrule
\end{tabular}%
}
\end{table}
We analyzed the network dynamics of nine different time periods within the March 24 - 28, 2020 Covid19 dataset, and visualized them based on their speed of retweeting. These types of graphs have been referred to as "retweet cascades" which describes how a social media network propagates information \cite{jin2017detection}.  Similar methods have been applied for visualizing rumor propogation by Jin et al. \cite{jin2017detection} We wanted to analyze how Covid19 retweeting behaves at different time points. We used published disaster retweeting times to serve as benchmarks for selecting time periods.  As a result, the graphs in \autoref{networkGraphs} are plotted by retweeting time of known benchmarks - the median time to retweet after an earthquake which implies rapid notification, the median time to retweet after a funnel cloud has been seen, all the way to a one-day or 24 hour time period. We did this to visualize a retweet cascade of fast to slow information propogation. We used median retweeting times published Spiro et al. \cite{spiro2012waiting} for the time it took users to retweet messages based on hazardous keywords like "Funnel Cloud", "Aftershock", and "Mudslide".  We also used the H7N9 reposting time which Zhang et al. \cite{zhang2017social} published of 3.7 hours.  

\begin{figure}[t]
\centering
\includegraphics[width=\columnwidth]{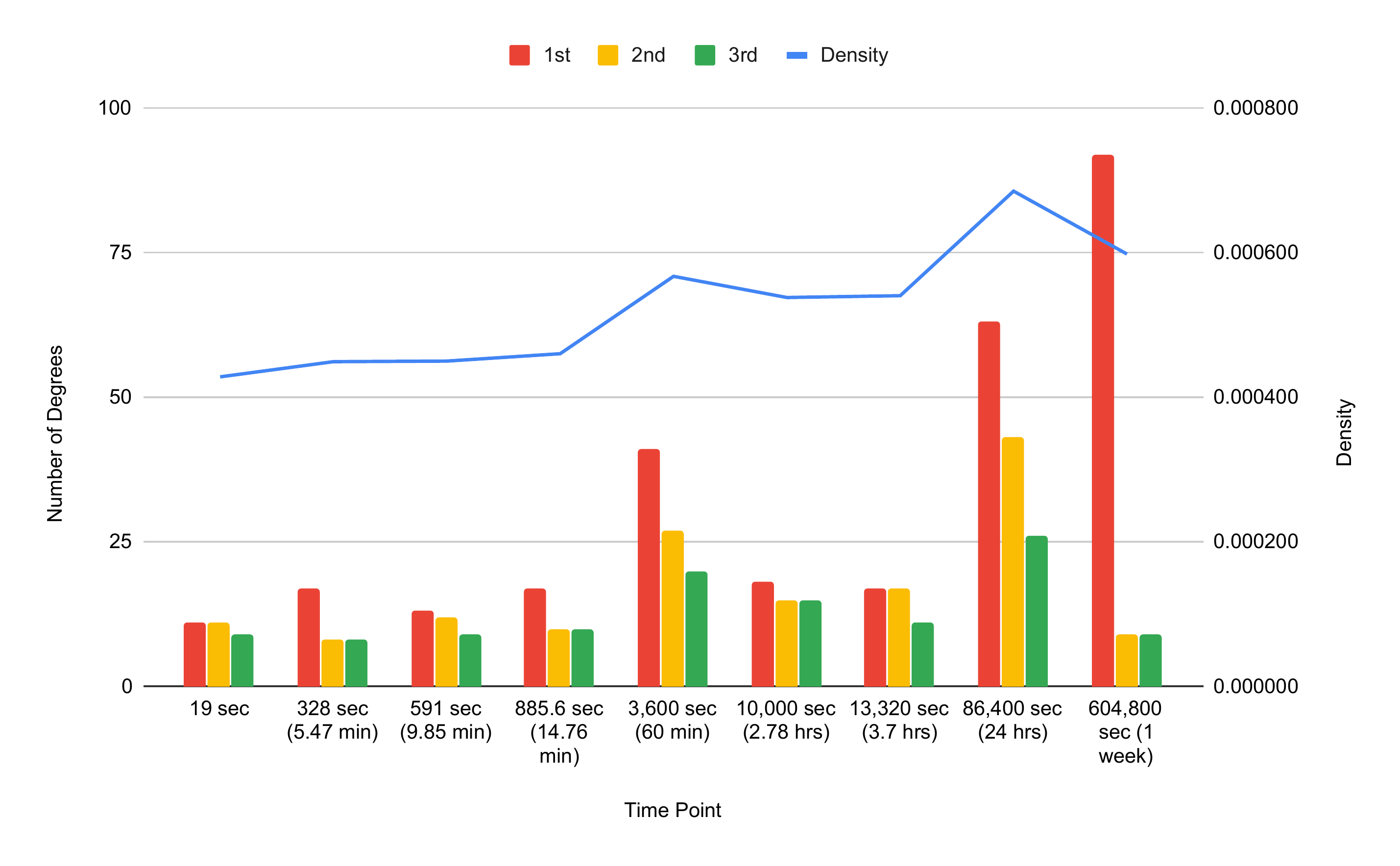}
\caption{Increasing Density and Degree for Top 3 Users}
\label{graph_stats}
\end{figure}

We generated a Directed Graph for each of the nine time periods, where the network consisted of a source which was the author of the tweet (User object, the id\_str) and a target which was the original retweeter shown in \autoref{graph_stats_table}. The goal was to analyze how connections change as the retweeting speed increases. The nine networks are visualized in \autoref{networkGraphs}. Graphs were plotted using networkx and drawn using the Kamada Kawai Layout\cite{kamada1989algorithm}, a force-directed algorithm. We modeled 700 users for each graph. We found that more nodes became too difficult to interpret. The size of the node indicates the number of degrees, or users that it is connected to. It can mean that the node has been retweeted by others several times. Or, it can also mean that the node itself has been retweeted by others several times. 

The density of each network increases over time shown in \autoref{networkGraphs} and \autoref{graph_stats}. Very rapid retweeters, in the time it takes to retweet after an earthquake, start off with a sparse network with a few nodes in the center being the focus of retweets in \autoref{8a}. By the time we reach \autoref{d}, the retweeted users are much more clustered in the center and there are more connections and activity. The top retweeted user in our median time network \autoref{g}, was a news network and tweeted \emph{"The team took less than a week to take the ventilator from the drawing board to working prototype, so that it can"}. By 24 hours out in \autoref{h}, we see a concentrated set of users being retweeted and by \autoref{i}, one account appears to dominate the space being retweeted 92 times. This account was retweeting the following message several times \emph{"She was doing \#chemotherapy couldn’t leave the house because of the threat of \#coronavirus so her line sisters..."}. In addition, the number of nodes generally decreased from 1278 in "earthquake" time to 1067 in one week, and the density also generally increased, shown in \autoref{graph_stats_table}. 

These retweet cascade graphs provide only an exploratory analysis. Network structures like these have been used to predict virality of messages, for example memes over time as the message is diffused across networks \cite{weng2013virality}. But, analyzing them further could enable 1) an improved understanding about how Covid19 information diffusion is different than other outbreaks, or global events, 2) How information is transmitted differently from region to region across the world, and 3) What users and messages are being concentrated on over time. This would support strategies to improve government communications, emergency messaging, dispelling medical rumors, and tailoring public health announcements.

\section{Limitations}
There are several limitations with this study.  First, our dataset is discontinuous and trends seen in \autoref{baxes1} and \autoref{baxes2} where there is an interruption in time should be taken with caution.  Although there appears to be a trend between one discrete time and another, without the missing data, it is impossible to confirm this as a trend.  As a result, it would be valuable to apply these techniques on a larger and continuous corpus without any time breaks. We aim to repeat the methods in this study on a longer continuous stream of Twitter data in the near future.  

Next, the corpus we analyzed was already pre-filtered with thirteen "track" terms from the Twitter Streaming API that focused the dataset towards healthcare related concerns. This may be the reason why the high level keywords extracted in the first round of analysis were consistently mentioned throughout the different stages of modeling. However, after review of similar papers indicated in \autoref{papers}, we found that despite having filtered the corpus on healthcare-related terms, topics still appear to be consistent with analyses where corpora were filtered on limited terms like "\#coronavirus". 

Third, the users and conversations in Twitter are not a direct representation of the U.S. or global population. The Pew Research Foundation found that only 22\% of American adults use Twitter \cite{perrin_anderson_2019} and that this group is different from the majority of U.S. adults, because they are on average younger, more likely to identify as Democrats, more highly educated and possess higher incomes \cite{wojcik_hughes_2020}.  The users were also not verified and should be considered as a possible mixture of human and bot accounts. 

Fourth, we reduced our corpus to remove retweets for the keyword and topic modeling anlayses since retweets can obscure the message by introducing virality and altering the perception of the information \cite{madrigal_2018}. As a result, this reduced the size of our corpus by nearly 77\% from 23,820,322 tweets to 5,506,223 tweets.  However, there appears to be variability in terms of consistent corpora sizes in the Twitter analysis literature both in \autoref{papers} and other health-related studies. For example, Karami \cite{karami2018characterizing} used 4.5 million tweets, Zhao \cite{zhao2011comparing} used 1,225,851 tweets, Hong\cite{hong2010empirical}used 1,992,758 tweets,  Surian \cite{surian2016characterizing} used 285,417 tweets, Alverez\cite{alvarez2016topic} used 101,522 tweets, and Lim \cite{lim2016twitter} used only 60,370 tweets. 

Fifth, our compute limitations prohibited us from analyzing a larger corpus for the UMAP, time-series, and network modeling. For the LDA models we leveraged the gensim MulticoreLDA model that allowed us to leverage multiprocessing across 20 workers. But for UMAP and the network modeling, we were constrained to use a CPU. However, as stated above, visualizing more than 700 nodes for our graph models was unintepretable. Applying our methods across the entire 23.8 million corpora for UMAP and the network models may yield more meaningful results. Sixth, we were only able to iterate over 15 different LDA models based on changing the number of topics, whereas Syed et al. \cite{syed2017full} iterated on 480 models to select coherent models. We believe that applying a manual gridsearch of the LDA parameters such as iterations, alpha, gamma threshold, chunksize, and number of passes would lead to a more diverse representation of LDA models and possibly more coherent topics.  

Seven, it was challenging to identify papers that analyzed Twitter networks according to their speed of retweets for public health emergencies and disease outbreaks.  Zhang et al. \cite{zhang2017social} points out that there are not enough studies of temporal measurement of public response to health emergencies. We were lucky to find papers by Zhang et al. \cite{zhang2017social} and Spiro et al. \cite{spiro2012waiting} who published on disaster waiting times. Chew et al. \cite{chew2010pandemics} and Szomszor et al. \cite{szomszor2010swineflu} have published about Twitter analysis in H1N1 and the Swine Flu, respectively. Chew analyzed the volume of H1N1 tweets and categorized different types of messages such as humor and concern. Szomszor correlated tweets with UK national surveillance data and Tang et al. \cite{tang2018tweeting} generated a semantic network of tweets on measles during the 2015 measles outbreak to understand keywords mentioned about news updates, public health, vaccines and politics. However, it was difficult to compare our findings against other disease outbreaks due to the lack of similar modeling and published retweet cascade times and network models. 

\section{Conclusion}
We answered five research questions about Covid19 tweets during March 24, 2020 - April 8, 2020. First, we found high-level trends that could be inferred from keyword analysis. Second, we found that live White House Coronavirus Briefings led to spikes in Topic 18 ("potus"). Third, using UMAP, we found strong local "clustering" of topics representing PPE, healthcare workers, and government concerns. UMAP allowed for an improved understanding of distinct topics generated by LDA. Fourth, we used retweets to calculate the speed of retweeting.  We found that the median retweeting time was 2.87 hours. Fifth, using directed graphs we plotted the networks of Covid19 retweeting communities from rapid to longer retweeting times. The density of each network increased over time as the number of nodes generally decreased. 

Lastly, we recommend trying all techniques indicated in \autoref{papers} to gain an overall understanding of Covid19 Twitter data. While applying multiple methods for an exploratory strategy, there is no technical guarantee that the same combination of five methods analyzed in this paper will yield insights on a different time period of data. As a result, researchers should attempt multiple techniques and draw on existing literature.

\section*{Acknowledgment}
The authors would like to acknowledge John Larson from Booz Allen Hamilton for his support and review of this article.

\newpage
\bibliographystyle{IEEEtranN} 

\bibliography{references}

\newpage
\onecolumn
\appendices

\section{Twitter Dataset in UTC Time}
\begin{table*}[ht]
\caption{Twitter Data Sets March 24, 2020 - April 8, 2020}
\label{datasets}
\resizebox{\textwidth}{!}{%
\begin{tabular}{@{}llllllll@{}}
\toprule
\textbf{Corpus} &
  \textbf{Time Start} &
  \textbf{Time End} &
  \textbf{Total Minutes} &
  \textbf{Size, GB} &
  \textbf{Total Tweets} &
  \textbf{No Retweets} &
  \textbf{Perc No Retweets} \\ \midrule
3/24/2020 & 2020-03-24 21:17:27+00:00 & 2020-03-24 22:00:48+00:00 & 44   & 1    & 132,658   & 27,374    & 20.64\% \\
3/25/2020 & 2020-03-25 14:45:12+00:00 & 2020-03-25 16:18:47+00:00 & 94   & 2    & 286,405   & 63,649    & 22.22\% \\
3/28/2020 & 2020-03-28 00:17:20+00:00 & 2020-03-28 02:01:08+00:00 & 105  & 2.3  & 317,498   & 61,933    & 19.51\% \\
3/30/2020 & 2020-03-30 12:55:38+00:00 & 2020-03-30 21:44:35+00:00 & 530  & 11.5 & 1,618,620 & 365,808   & 22.60\% \\
3/31/2020 & 2020-03-30 21:47:53+00:00 & 2020-03-31 13:15:36+00:00 & 929  & 20.3 & 2,802,069 & 576,741   & 20.58\% \\
4/4/2020  & 2020-04-03 00:29:11+00:00 & 2020-04-04 22:05:12+00:00 & 2737 & 56.2 & 7,755,704 & 1,795,912 & 23.16\% \\
4/5/2020  & 2020-04-05 20:41:43+00:00 & 2020-04-07 15:07:11+00:00 & 2547 & 49.4 & 6,810,216 & 1,599,455 & 23.49\% \\
4/8/2020  & 2020-04-08 13:54:33+0000  & 2020-04-09 14:30:54+0000  & 1477 & 30.4 & 4,107,152 & 1,015,351 & 24.72\% \\
\textbf{Total} &
  \textbf{} &
  \textbf{} &
  \textbf{8463} &
  \textbf{173.1} &
  \textbf{23,830,322} &
  \textbf{5,506,223} &
  \textbf{23.11\%} \\ \bottomrule
\end{tabular}%
}
\end{table*}

\begin{table*}[ht!]
\centering
\caption{Keyword Raw Counts}
\resizebox{\textwidth}{!}{%
\begin{tabular}{@{}llllllllllllll@{}}
\toprule
\textbf{Corpus} &
  \textbf{bed} &
  \textbf{hospital} &
  \textbf{mask} &
  \textbf{icu} &
  \textbf{help} &
  \textbf{nurse} &
  \textbf{doctors} &
  \textbf{vent} &
  \textbf{test\_pos} &
  \textbf{serious\_cond} &
  \textbf{exposure} &
  \textbf{cough} &
  \textbf{fever} \\ \midrule
3/24/2020 & 147    & 1,323   & 1,685   & 139    & 114    & 215    & 372    & 1,143   & 28    & 1   & 11    & 18  & 1   \\
3/25/2020 & 293    & 3,104   & 3,641   & 265    & 299    & 352    & 758    & 2,321   & 122   & 4   & 26    & 35  & 10  \\
3/28/2020 & 191    & 3,218   & 3,607   & 180    & 248    & 504    & 891    & 4,073   & 101   & 2   & 19    & 19  & 3   \\
3/30/2020 & 1,475  & 21,707  & 28,512  & 1,225  & 1,742  & 3,708  & 6,895  & 13,190  & 589   & 13  & 114   & 157 & 23  \\
3/31/2020 & 1,959  & 28,495  & 67,703  & 1,344  & 3,416  & 5,233  & 9,671  & 16,717  & 948   & 13  & 141   & 459 & 137 \\
4/2/2020  & 5,652  & 80,495  & 231,185 & 4,034  & 8,661  & 16,823 & 28,603 & 64,112  & 2,228 & 48  & 525   & 977 & 122 \\
4/5/2020  & 5,648  & 81,025  & 159,915 & 6,350  & 7,741  & 14,767 & 27,341 & 45,614  & 2,612 & 36  & 445   & 786 & 133 \\
Total     & 15,365 & 219,367 & 496,248 & 13,537 & 22,221 & 41,602 & 74,531 & 147,170 & 6,628 & 117 & 1,281 & 688 & 429 \\ \bottomrule
\end{tabular}%
}
\label{appendix_rawcounts}
\end{table*}

\section{Topic Modeling Implementation Details}

For the LDA topic modeling, we used the gensim Python library \cite{rehurek_lrec, rehurek_2014}. It provides four different coherence metrics. We used the "c\_v" metric for coherence developed by Roder\cite{roder2015exploring}. Coherence metrics are used to rate the quality and human interpretability of a topic generated. All models were run with the default parameters using a LdaMulticore model parallel computing on 20 workers, default gamma threshhold of 0.001, chunksize of 10,000, 100 iterations, 2 passes.

\newpage
\section{Live Press Briefings and Topic Time Series}

\begin{figure*}[ht]
\centering
\includegraphics[width=1.0\textwidth]{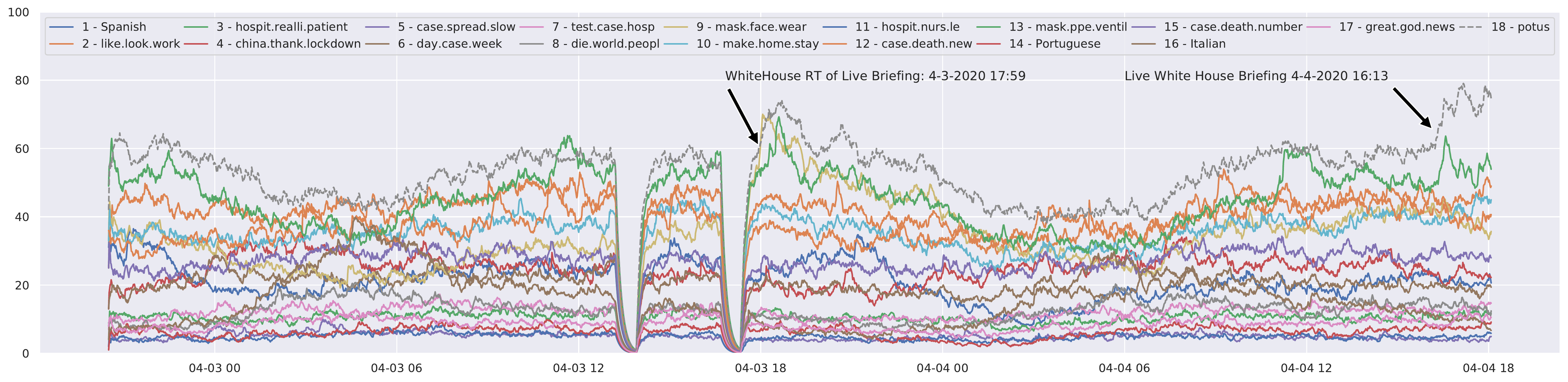}
\caption{April 3 8:29 PM EST to April 4 6:05 PM EST Topics Time Series}
\label{WHbriefingapr3}
\end{figure*}

Note - Sudden decreases in \autoref{WHbriefingapr3} signal may be due to temporary internet disconnection.

\begin{figure*}[ht!]
\centering
\includegraphics[width=1.0\textwidth]{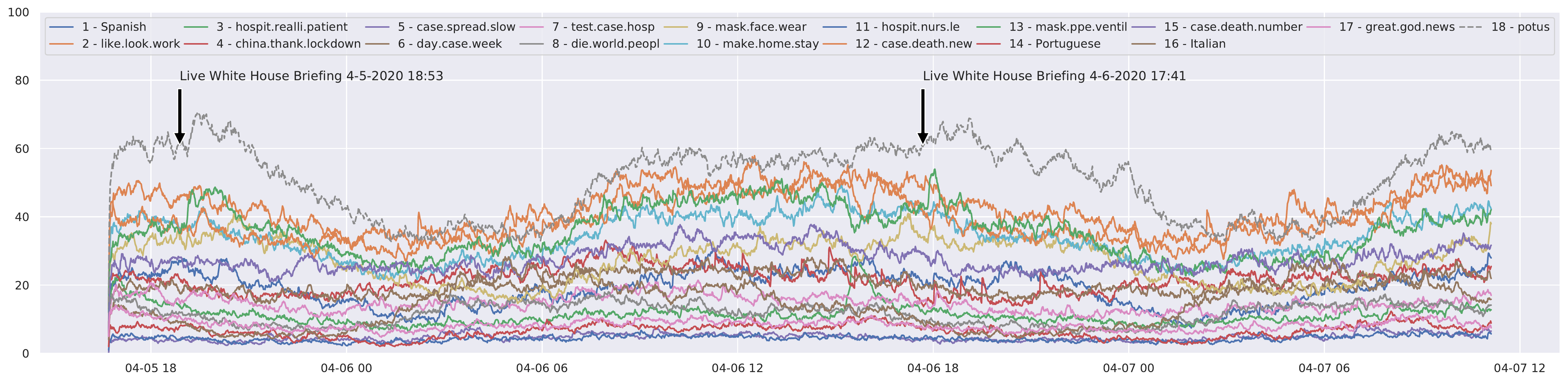}
\caption{April 5 4:41 PM EST to  April 7 11:07 AM EST Topics Time Series}
\label{WHbriefingapr5}
\end{figure*}

\begin{figure*}[ht!]
\centering
\includegraphics[width=1.0\textwidth]{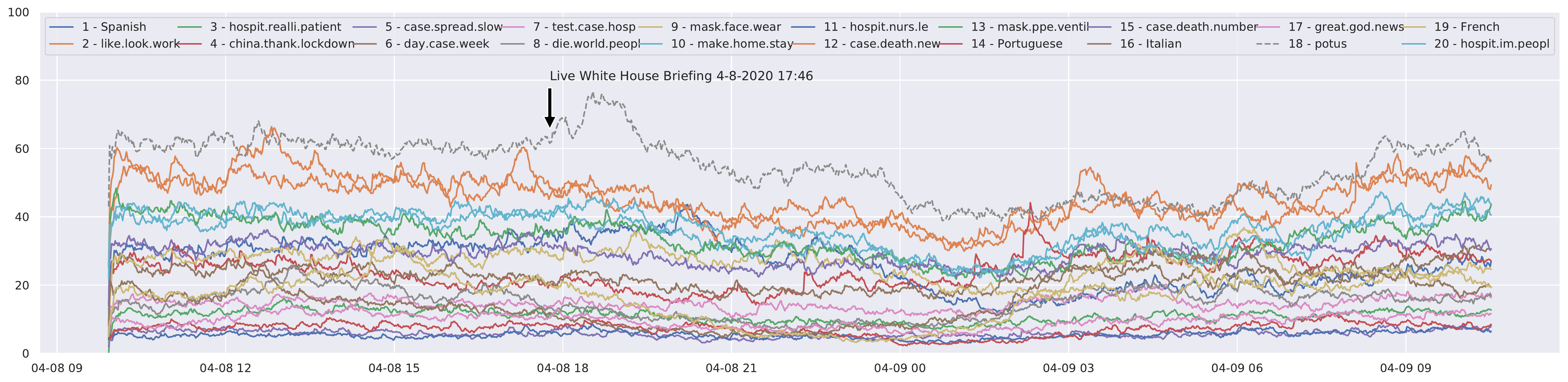}
\caption{April 8 9:54 AM EST to April 9 10:30 AM EST Topics Time Series}
\label{WHbriefingapr8}
\end{figure*}

\newpage
\section{Change Point Detection Time Series}
Models were calculated using the ruptures Python package. We also applied exponential weighted moving average using the ewm pandas function. We applied a span of 5 for March 24, 2020 and a span of 20 for April 3 - 4 datasets, April 5 - 6 datasets, and April 8 - 9 datasets. Our parameters for binary segmentation included selecting the "l2" model to fit the points for Topic 18, using 10 n\_bkps (breakpoints).

\begin{figure*}[ht!]
\begin{subfigure}{\textwidth}
\centering
  \includegraphics[width=0.75\linewidth]{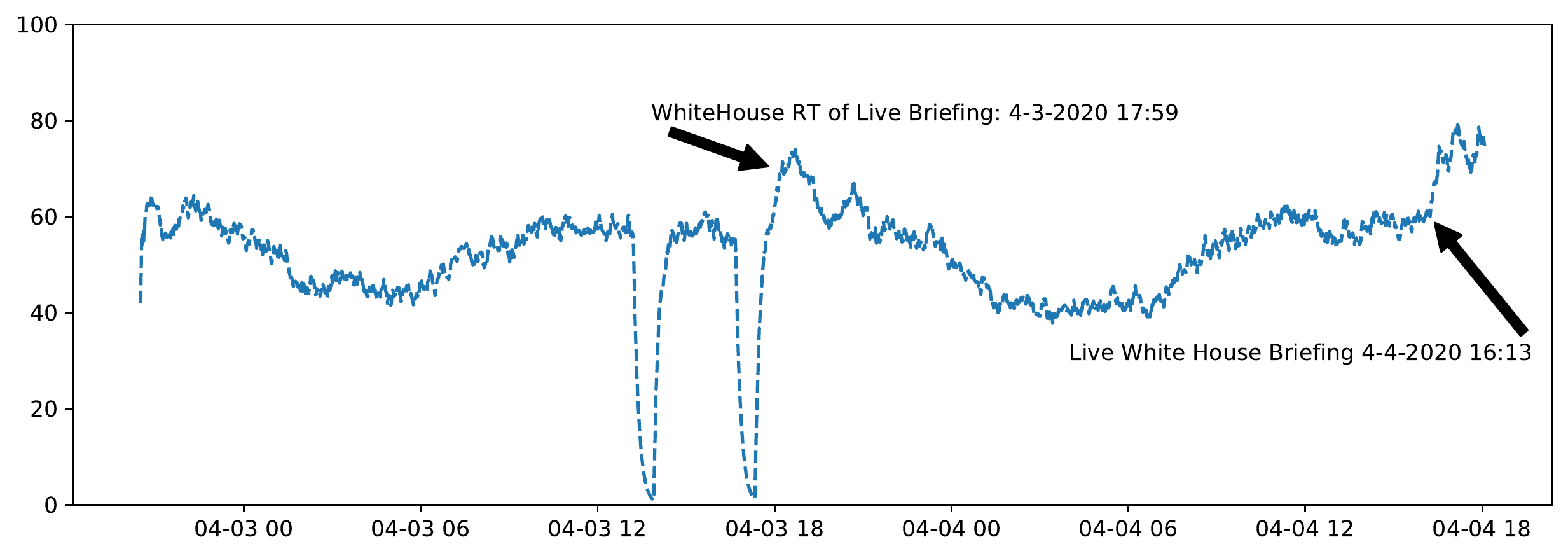}
  \label{a}
\end{subfigure}%
\newline
\begin{subfigure}{\textwidth}
\centering
  \includegraphics[width=0.75\linewidth]{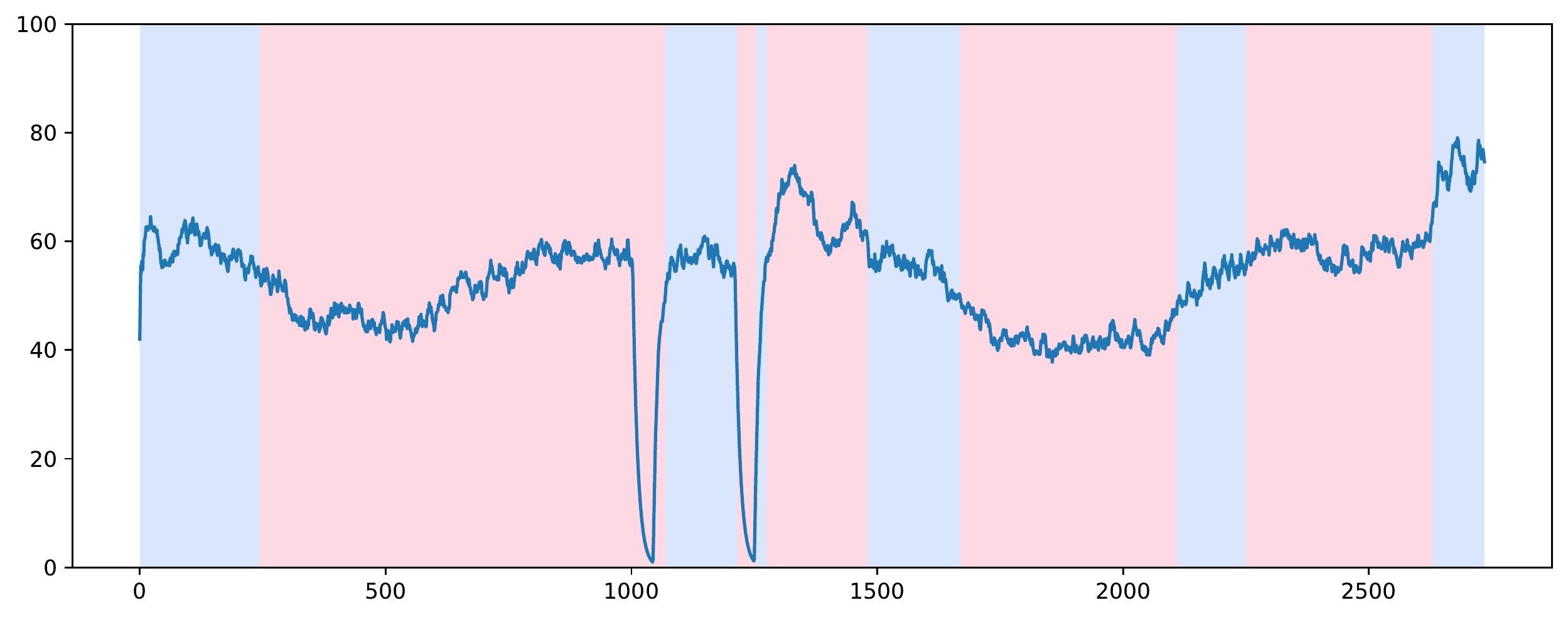}
  \label{b}
\end{subfigure}
\caption{Change Point Detection using Binary Segmentation for April 3 - 4, 2020}
\label{cpd_apr34}
\end{figure*}

\begin{figure*}[ht!]
\begin{subfigure}{\textwidth}
\centering
  \includegraphics[width=0.75\linewidth]{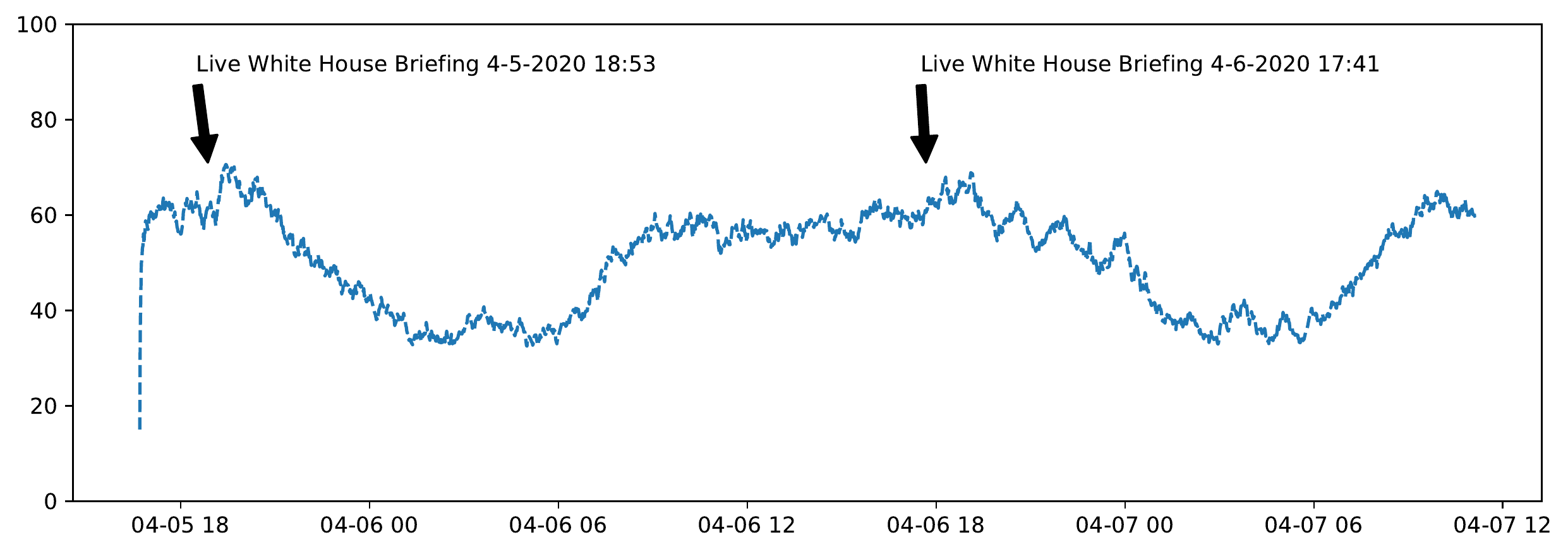}
  \label{a}
\end{subfigure}%
\newline
\begin{subfigure}{\textwidth}
\centering
  \includegraphics[width=0.75\linewidth]{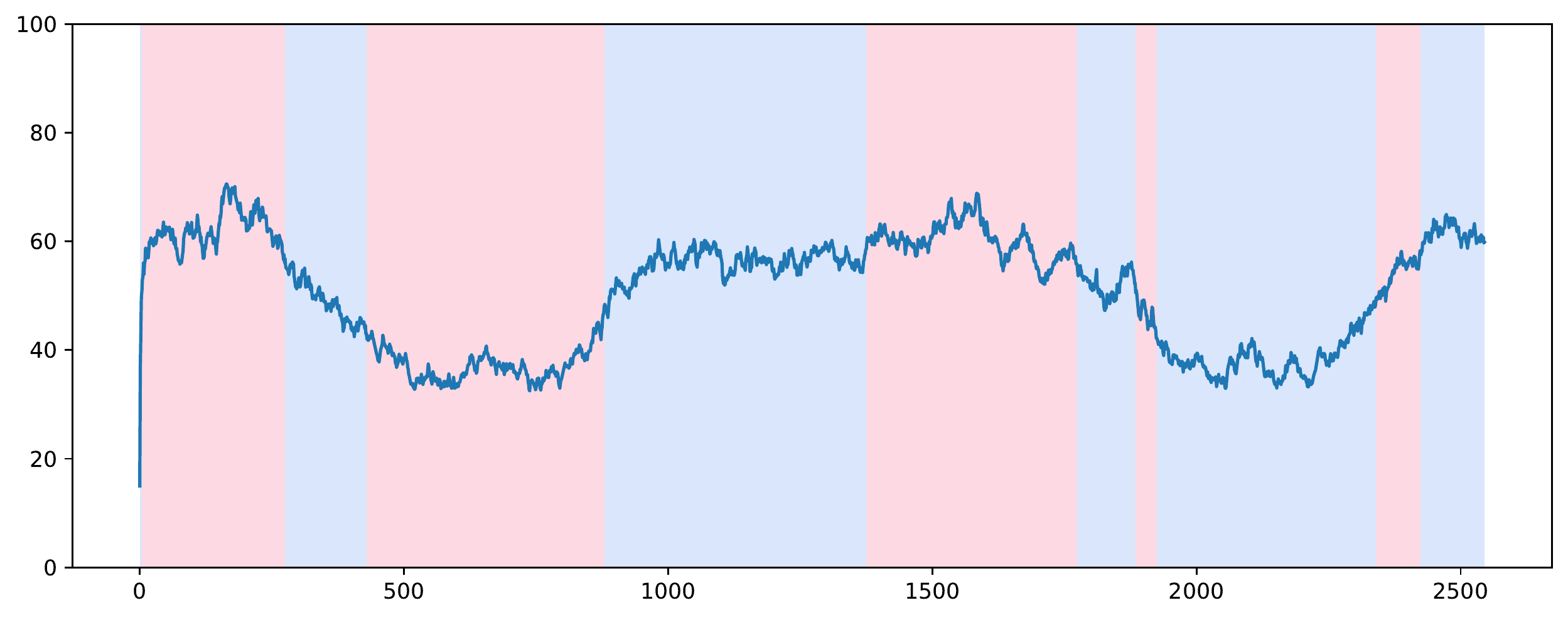}
  \label{b}
\end{subfigure}
\caption{Change Point Detection using Binary Segmentation for April 5, 2020}
\label{cpd_apr5}
\end{figure*}

\begin{figure*}[ht!]
\begin{subfigure}{\textwidth}
\centering
  \includegraphics[width=0.75\linewidth]{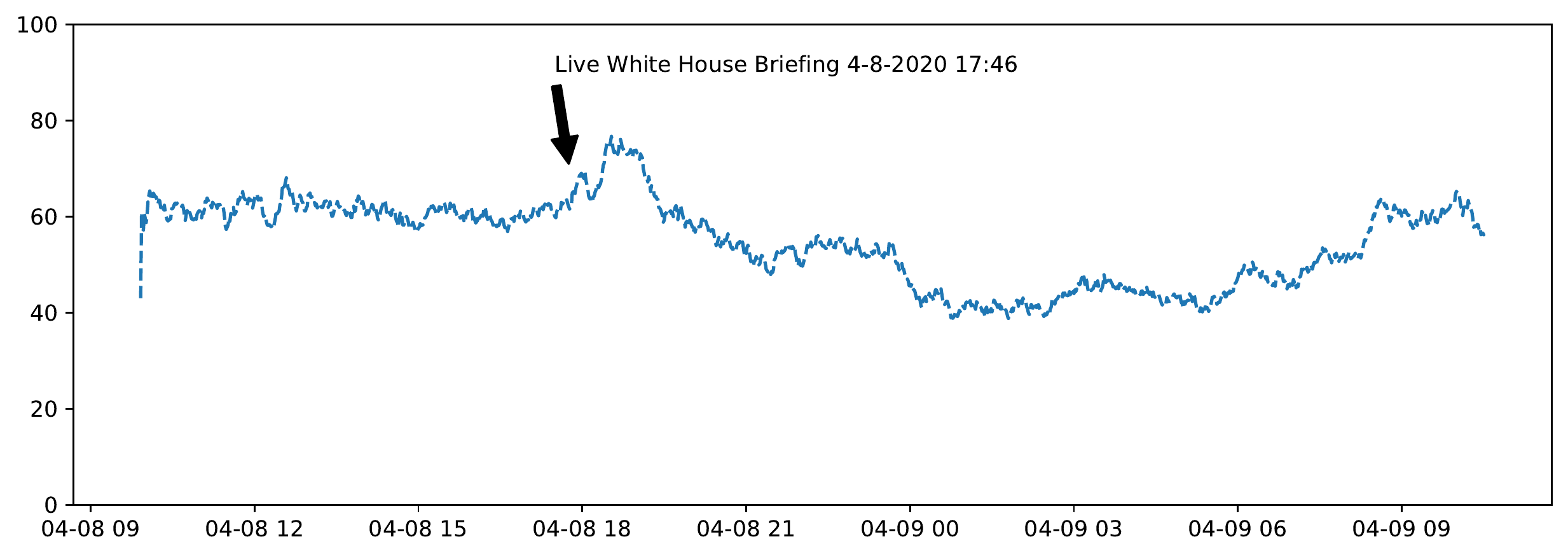}
  \label{a}
\end{subfigure}%
\newline
\begin{subfigure}{\textwidth}
\centering
  \includegraphics[width=0.75\linewidth]{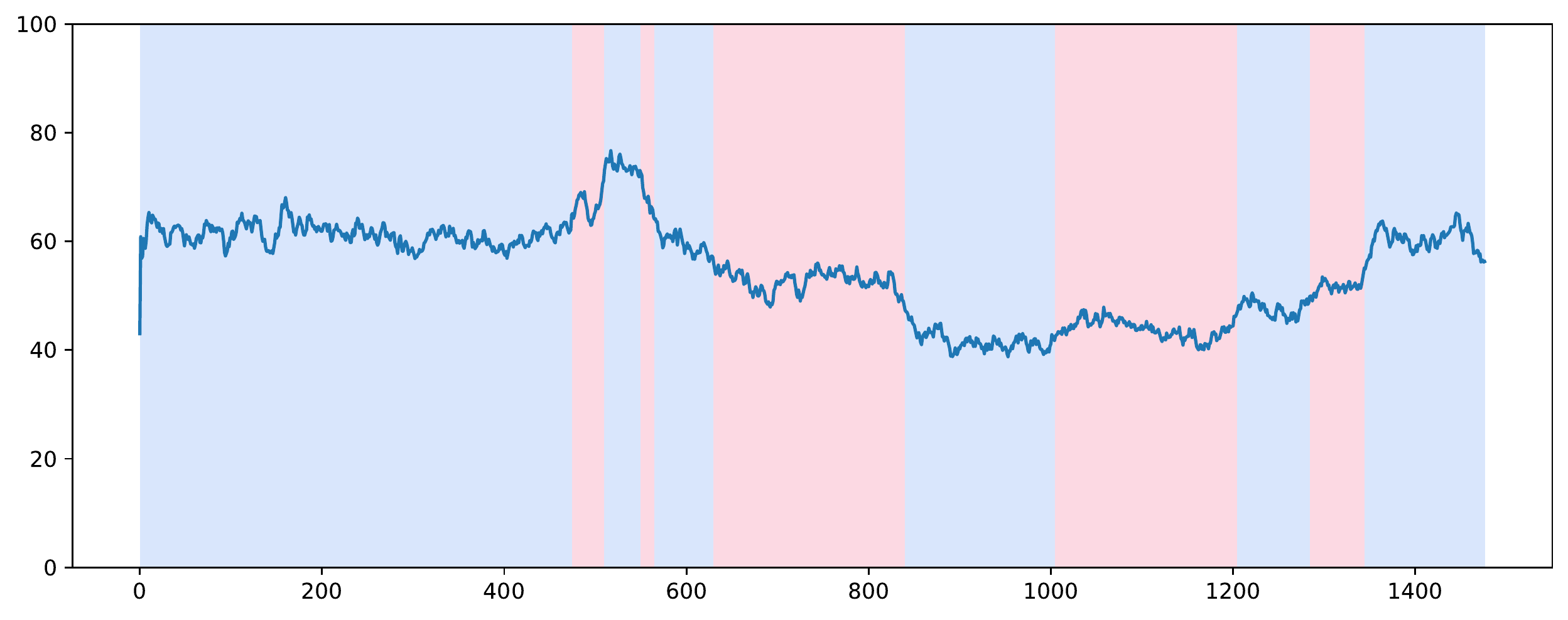}
  \label{b}
\end{subfigure}
\caption{Change Point Detection using Binary Segmentation for April 8, 2020}
\label{cpd_apr8}
\end{figure*}

\newpage
\section{TF-IDF Frequencies of Tweets Rapidly Retweeted}

\end{document}